\documentclass[11pt]{combine}
\usepackage{amsmath,amsfonts,amssymb,amsthm,epsfig, graphicx, hyperref}
\usepackage[dvipsnames,table,dvipsnames*, svgnames*, hyperref]{xcolor}
\usepackage{natbib,amsmath,amssymb,amsthm,graphicx,setspace,paralist,booktabs,rotating,subcaption,float,color}
\usepackage{multirow}
\usepackage{fancyhdr}
\usepackage{bibunits}
\usepackage[small]{titlesec}

\newtheorem{theorem}{Theorem}
\newtheorem{corollary}{Corollary}
\newtheorem{proposition}{Proposition}

\newtheorem{lemma}{Lemma}
{
\theoremstyle{definition}
\newtheorem{definition}{Definition}
\newtheorem{example}{Example}
\newtheorem{remark}{Remark}
}
\newcommand{\beq}{\begin{equation}}
\newcommand{\eeq}{\end{equation}}
\newcommand{\beas}{\begin{align*}}
\newcommand{\eeas}{\end{align*}}
\newcommand{\bea}{\begin{align}}
\newcommand{\eea}{\end{align}}
\newcommand{\bei}{\begin{itemize}}
\newcommand{\eei}{\end{itemize}}
\newcommand{\ben}{\begin{enumerate}}
\newcommand{\een}{\end{enumerate}}
\newcommand{\bet}{\begin{theorem}}
\newcommand{\eet}{\end{theorem}}
\newcommand{\bel}{\begin{lemma}}
\newcommand{\eel}{\end{lemma}}
\newcommand{\bep}{\begin{proposition}}
\newcommand{\eep}{\end{proposition}}
\newtheorem{proc}{Procedure}
\newcommand{\bed}{\begin{definition}}
\newcommand{\eed}{\end{definition}}
\newcommand{\bec}{\begin{corollary}}
\newcommand{\eec}{\end{corollary}}
\newcommand{\bex}{\begin{example}}
\newcommand{\eex}{\end{example}}

\newcommand{\Ome}{\bold{\Omega}}
\newcommand{\Sig}{\bold{\Sigma}}

\newcommand{\R}{\mathbb{R}}
\newcommand{\E}{\mathbb{E}}

\newcommand{\HH}{\mathcal{H}}

\newcommand{\argmin}{\mathop{\rm arg\min}}

\numberwithin{equation}{section}

\def\bx{\bold{x}}

\def\limsup{\mathop{\overline{\rm lim}}}
\def\liminf{\mathop{\underline{\rm lim}}}

\begin{document}


\title{\scshape Global and Simultaneous Hypothesis Testing for High-Dimensional Logistic Regression Models}
\author{Rong Ma$^1$, T. Tony Cai$^2$ and Hongzhe Li$^1$ \\
Department of Biostatistics, Epidemiology and Informatics$^1$\\
Department of Statistics$^2$\\
University of Pennsylvania\\
Philadelphia, PA 19104}
\date{}
\maketitle
\thispagestyle{empty}

	\begin{abstract}
High-dimensional logistic regression is  widely used in analyzing data with binary outcomes.  In this paper, global testing and large-scale multiple testing for the regression coefficients are considered in both single- and two-regression settings.  A test statistic for testing the global null hypothesis is constructed using a generalized low-dimensional projection for bias correction and  its asymptotic null distribution is derived.   A lower bound for the global testing is established, which shows that the proposed test is asymptotically minimax optimal over some sparsity range. For testing the individual coefficients simultaneously, multiple testing procedures are proposed and shown to control the false discovery rate (FDR) and falsely discovered variables (FDV) asymptotically.   Simulation studies are carried out to examine the numerical performance of the proposed tests and their superiority over existing methods. The testing procedures are also illustrated by analyzing a data set of a metabolomics study that  investigates the association between fecal  metabolites and pediatric Crohn's disease and the effects of treatment on such associations. 
	\bigskip
	
	\noindent\emph{KEY WORDS}:  False discovery rate; Global testing;  Large-scale multiple testing; Minimax lower bound.
\end{abstract}

\section{INTRODUCTION}

Logistic regression models have been applied widely in  genetics, finance, and business analytics.  In many modern applications,  the number of covariates of interest usually grows with, and sometimes far exceeds, the number of observed samples. In such high-dimensional settings, statistical problems such as estimation, hypothesis testing, and construction of confidence intervals become much more challenging than those in the classical low-dimensional  settings. The increasing technical difficulties usually emerge from the non-asymptotic analysis of both statistical models and the corresponding computational algorithms. 

In this paper, we consider testing for  high-dimensional logistic regression model:
\beq  \label{logit.reg}
\log\bigg(\frac{\pi_i}{1-\pi_i}\bigg) = X_i^\top\beta,\quad\quad\text{for } i=1,...,n.
\eeq
where $\beta \in \R^p$ is the vector of regression coefficients. The observations are i.i.d. samples $Z_i = (y_i, X_i)$ for $i=1,..,n$, and we assume $y_i | X_i \sim \text{Bernoulli}(\pi_i)$ independently for each $i = 1,...,n.$ 

\subsection{Global and Simultaneous Hypothesis Testing}

It is important in high-dimensional logistic regression to determine 1) whether there are any associations between the covariates and the outcome and, if yes, 2) which covariates are associated with the outcome. The first  question can be formulated as testing the global null hypothesis $H_0: \, \beta=0$; and the second question can be considered as simultaneously testing the null hypotheses  $H_{0,i}:\, \beta_i=0$ for $i=1,...,p$.  Besides such single logistic regression problems,   hypothesis testing involving two logistic regression models with regression coefficients $\beta^{(1)}$ and $\beta^{(2)}$ in $\R^p$ is also important.  Specifically, one is interested in testing the global null hypothesis $H_{0}:\,  \beta^{(1)}=\beta^{(2)}$, or identifying the differentially associated covariates through simultaneously testing the null hypotheses $H_{0,i}:\,  \beta_i^{(1)}=\beta_i^{(2)}$ for each $i=1,...,p$. 

Estimation for high-dimensional logistic regression has been studied extensively. \cite{van2008high} considered high-dimensional generalized linear models (GLMs) with Lipschitz loss functions, and proved a non-asymptotic oracle inequality for the empirical risk minimizer with the
Lasso penalty. \cite{meier2008group} studied the group Lasso for logistic regression and proposed an efficient algorithm that leads to statistically consistent estimates. \cite{negahban2010unified} obtained the rate of convergence for the $\ell_1$-regularized maximum likelihood estimator under GLMs using restricted strong convexity property. \cite{bach2010self} extended tools from the convex optimization literature, namely self-concordant functions, to provide interesting extensions of theoretical results for the square loss to the logistic loss. \cite{plan2013robust} connected sparse logistic regression to one-bit compressed sensing and developed a unified theory for signal estimation with noisy observations.

In contrast, hypothesis testing and confidence intervals for high-dimensional logistic regression have only been recently addressed. \cite{van2014asymptotically} considered constructing confidence intervals and statistical tests for single or low-dimensional components of the regression coefficients in high-dimensional GLMs. \cite{mukherjee2015hypothesis} studied the detection boundary for minimax hypothesis testing in high-dimensional sparse binary regression models when the design matrix is sparse. \cite{belloni2016post} considered estimating and constructing the confidence regions for a regression coefficient of primary interest in GLMs. More recently, \cite{sur2017likelihood} and \cite{sur2019modern} considered the likelihood ratio test for high-dimensional logistic regression under the setting that $p/n\to \kappa$ for some constant $\kappa<1/2$, and showed that the asymptotic null distribution of  the log-likelihood ratio statistic is a rescaled $\chi^2$ distribution.  \cite{cai2017differential} proposed a global test and a multiple testing procedure for differential networks against sparse alternatives under the Markov random field model. Nevertheless, the problems of global testing and large-scale simultaneous testing for high-dimensional logistic regression models with $p\gtrsim n$ remain unsolved.

In this paper, we first consider global and multiple testing for a single high-dimensional logistic regression model.  The global test statistic is constructed as the maximum of squared standardized statistics for individual coefficients, which are based on a two-step standardization procedure. The first step is to correct the bias of the logistic Lasso estimator using a generalized low-dimensional projection (LDP) method, and the second step is to normalize the resulting nearly unbiased estimators by their estimated standard errors. We show that the asymptotic null distribution of the test statistic is a Gumbel distribution and that the resulting test is minimax optimal under the Gaussian design by establishing the minimax separation distance between the null space and alternative space.  For large-scale multiple testing, data-driven testing procedures are proposed and shown to control the false discovery rate (FDR) and  falsely discovered variables (FDV) asymptotically. The framework for  testing  for single logistic regression is then extended to the setting of testing two  logistic regression models.

The main contributions of the present paper are threefold.
\begin{enumerate}
	\item We propose novel procedures for  both the global testing and large-scale simultaneous testing for high dimensional logistic regressions. The dimension $p$ is allowed to be much larger than the sample size $n$. Specifically, we require $\log p=O(n^{c_1})$ for the global test and $p=O(n^{c_2})$ for the multiple testing procedure, with some constant $c_1,c_2>0$. For the global alternatives characterized by the $\ell_\infty$ norm of the regression coefficients, the global test is shown to be minimax rate optimal with the optimal separation distance of order $\sqrt{\log p/n}$.
	\item Following similar ideas in \cite{ren2016asymptotic} and \cite{cai2017differential}, our construction of the test statistics depends on a generalized version of the LDP method for bias correction.
	The original LDP method \citep{zhang2014confidence} relies on the linearity between the covariates and outcome variable. For logistic regression, the generalized approach first finds a linearization of the regression function, and the weighted LDP is then applied. Besides its usefulness in logistic regression, the generalized LDP method is flexible and can be applied to other nonlinear regression problems (see Section 7 for a detailed discussion). 
	\item The minimax lower bound is obtained for the global hypothesis testing under the Gaussian design. The lower bound depends on the calculation of the $\chi^2$-divergence between two logistic regression models. To the best of our knowledge, this is the first lower bound result for high-dimensional logistic regression under the Gaussian design.
\end{enumerate}

\subsection{Other Related Work}

We should note that a different but related problem, namely  inference for high-dimensional linear regression, has been well studied in the literature.  \cite{zhang2014confidence}, \cite{van2014asymptotically} and \cite{javanmard2014confidence,javanmard2014hypothesis} considered confidence intervals and testing for low-dimensional parameters of the high-dimensional linear regression model and developed methods based on a two-stage debiased estimator that corrects the bias introduced at the first stage due to regularization. \cite{cai2017confidence} studied minimaxity and adaptivity of confidence intervals for general linear functionals of the regression vector.

The problems of global testing and large-scale simultaneous testing for high-dimensional linear regression have been studied by \cite{liu2014hypothesis}, \cite{ingster2010detection} and more recently by \cite{xia2018two} and \cite{javanmard2019false}. However, due to the nonlinearity and the binary outcome, the approaches used in these works cannot be directly applied to logistic regression problems. 
In the Markov random field setting, \cite{ren2016asymptotic} and \cite{cai2017differential} constructed pivotal/test statistics based on the debiased LDP estimators for node-wise logistic regressions with binary covariates. However,  the results for sparse high-dimensional logistic regression models with general continuous covariates remain unknown.

Other related problems include joint testing and false discovery rate control for high-dimensional multivariate regression \citep{xia2018joint} and  testing for high-dimensional precision matrices and Gaussian graphical models \citep{liu2013gaussian,xia2015testing}, where the inverse regression approach  and de-biasing were carried out in the construction of the test statistics.  Such statistics were then used for testing the global null with extreme value type asymptotic null distributions or to perform multiple testing  that controls the false discovery rate. 

\subsection{Organization of the Paper and Notations}

The rest of the paper is organized as follows. In Section 2, we propose the global test and establish its optimality. Some comparisons with existing works are made in detail. In Section 3, we present the multiple testing procedures and show that they control the FDR/FDP or FDV/FWER asymptotically. The framework is extended to the two-sample setting in Section 4. In Section 5, the numerical performance of the proposed tests are evaluated through extensive simulations. In Section 6, the methods are illustrated by an analysis of a metabolomics study. Further extensions and related problems are discussed in Section 7. In Section 8, some of the main theorems are proved. The proofs of other theorems as well as technical lemmas, and some further discussions are collected in the online Supplementary Materials.

Throughout our paper, for a vector $\bold{a} = (a_1,...,a_n)^\top \in \mathbb{R}^{n}$, we define the $\ell_p$ norm $\| \bold{a} \|_p = \big(\sum_{i=1}^n a_i^p\big)^{1/p}$, and the $\ell_\infty$ norm $\| \bold{a}\|_{\infty} = \max_{1\le j\le n}  |a_{i}|$. $\bold{a}_{-j}\in \R^{n-1}$ stands for the subvector of $\bold{a}$ without the $j$ the component. We denote $\text{diag}(a_1,...,a_n)$ as the $n\times n$ diagonal matrix whose diagonal entries are $a_1,...,a_n$. For a matrix $A\in \R^{p\times q}$, $\lambda_i(A)$ stands for the $i$-th largest singular value of $A$ and $\lambda_{\max}(A)  = \lambda_1(A)$, $\lambda_{\min}(A) = \lambda_{p \wedge q} (A)$. For a smooth function $f(x)$ defined on $\R$, we denote $\dot{f}(x) = d f(x)/dx$ and $\ddot{f}(x) = d^2 f(x)/dx^2$.  Furthermore, for sequences $\{a_n\}$ and $\{b_n\}$, we write $a_n = o(b_n)$ if $\lim_{n} a_n/b_n =0$, and write $a_n = O(b_n)$, $a_n\lesssim b_n$ or $b_n \gtrsim a_n$ if there exists a constant $C$ such that $a_n \le Cb_n$ for all $n$. We write $a_n\asymp b_n$ if $a_n \lesssim b_n$ and $a_n\gtrsim b_n$. For a set $A$, we denote $|A|$ as its cardinality. Lastly, $C, C_0, C_1,...$ are constants that may vary from place to place.

\section{GLOBAL HYPOTHESIS TESTING}

In this section, we consider testing the global null hypotheses 
\[
H_0: \beta=0 \quad \quad \text{vs.}  \quad\quad H_1:  \beta\ne 0,
\]
under the logistic regression model with random designs. The global testing problem corresponds to the detection of any associations between the covariates and the outcome.


Our construction of  the global testing procedure begins with a bias-corrected estimator built upon a regularized estimator such as the $\ell_1$-regularized M-estimator. For high-dimensional logistic regression, the $\ell_1$-regularized M-estimator is defined as
\beq \label{lasso.beta}
\hat{\beta} = \argmin_{\beta} \bigg\{\frac{1}{n}\sum_{i=1}^n \bigg[-y_i\beta^\top X_i+\log(1+e^{\beta^\top X_i}) \bigg]+\lambda \|\beta\|_1\bigg\},
\eeq
which is the minimizer of a penalized log-likelihood function.
\cite{negahban2010unified} showed that, when $X_i$ are i.i.d. sub-gaussian, under some mild regularity conditions, standard high-dimensional estimation error bounds for $\hat{\beta}$ under the $\ell_1$ or $\ell_2$ norm can be obtained by choosing $\lambda \asymp \sqrt{{\log p}/{n}}.$
Once we obtain the initial estimator $\hat{\beta}$, our next step is to correct the bias of $\hat{\beta}$.

For technical reasons, we split the samples so that the initial estimation step and the bias correction step are conducted on separate and independent datasets. Without loss of generality, we assume there are $2n$ samples, divided into two subsets $\mathcal{D}_1$ and $\mathcal{D}_2$, each with $n$ independent samples. The initial estimator $\hat{\beta}$ is obtained from $\mathcal{D}_1$. In the following, we construct a nearly unbiased estimator $\check{\beta}$ based on $\hat{\beta}$ and the samples from $\mathcal{D}_2$, using the generalized LDP approach. Throughout the paper, the samples $Z_i = (X_i,Y_i)$, $i=1,...,n$, are from $\mathcal{D}_2$, which are independent of $\hat{\beta}$. 
We would like to emphasize that the sample splitting procedure is only used to simplify our theoretical analysis, which does not  make it a restriction for practical applications. Numerically, as our simulations in Section 5 show, sample splitting is in fact not needed in order for  our methods perform well (see further discussions in Section 7).

\subsection{Construction of the Test Statistic via Generalized Low-Dimensional Projection}

Let $X$ be the design matrix whose $i$-th row is $X_i$.  We rewrite the logistic regression model defined by (\ref{logit.reg})  as 
\beq \label{reg}
y_i = f(\beta^\top X_i)+\epsilon_i
\eeq
where $f(u) = {e^u}/({1+e^u})$ and $\epsilon_i$ is error term. 
To correct the bias of the initial estimator $\hat{\beta}$, we consider the Taylor expansion of $f(u_i)$ at $\hat{u}_i$ for $u_i  =\beta^\top X_i$ and $\hat{u}_i = \hat{\beta}^\top X_i$
\[
f(u_i)=f(\hat{u}_i) +\dot{f}(\hat{u}_i)(u_i-\hat{u}_i)+Re_i
\]
where $Re_i$ is the reminder term. Plug this into the regression model (\ref{reg}), we have
\beq \label{reg.2}
y_i - f(\hat{u}_i)+\dot{f}(\hat{u}_i)X_i^\top \hat{\beta} = \dot{f}(\hat{u}_i)X_i^\top \beta+(Re_i+\epsilon_i).
\eeq
By rewriting the logistic regression model as (\ref{reg.2}), we can treat $y_i - f(\hat{u}_i)+\dot{f}(\hat{u}_i)X_i^\top \hat{\beta}$ on the left hand side as the new response variable, whereas $ \dot{f}(\hat{u}_i)X_i$ as the new covariates and $Re_i+\epsilon_i$ as the noise. Consequently, $\beta$ can be considered as the regression coefficient of this approximate linear model.

The bias-corrected estimator, or, the generalized LDP estimator $\check{\beta}$  is defined as
\beq \label{debias.beta}
\check{\beta}_j = \hat{\beta}_j + \frac{\sum_{i=1}^n v_{ij}(y_i-f(\hat{\beta}^\top X_i))}{\sum_{i=1}^nv_{ij}\dot{f}(\hat{\beta}^\top X_i)X_{ij}}, \quad j=1,...,p,
\eeq
where $X_{ij}$ is the $j$-th component of $X_i$ and $v_j = (v_{1j},v_{2j},..., v_{nj})^\top $ is the score vector that will be determined carefully \citep{ren2016asymptotic, cai2017differential}.
More specifically, we define the weighted inner product $\langle\cdot,\cdot\rangle_{n}$ for any $a,b\in \R^n$ as $\langle a,b \rangle_{n}  =  \sum_{i=1}^n \dot{f}(\hat{u}_i)a_ib_i$,
and denote $\langle\cdot,\cdot\rangle$ as the ordinary inner product defined in Euclidean space. 
Combining (\ref{reg.2}) and (\ref{debias.beta}), we can write
\beq \label{decomp1}
\check{\beta}_j - \beta_j = \frac{\langle v_j,\bold{\epsilon}\rangle}{\langle v_j,\bold{x}_j \rangle_n}+\frac{\langle v_j, Re \rangle}{\langle v_j,\bold{x}_j \rangle_n} - \frac{\langle v_j,\bold{h}_{-j}\rangle_n}{\langle v_j,\bold{x}_j \rangle_n},
\eeq
where $\bx_j\in \R^n$ denote the $j$-th column of $X$, $\bold{h}_{-j} = X_{-j}(\hat{\beta}_{-j}-\beta_{-j})$ where $X_{-j}\in \R^{n}\times \R^{p-1}$ is the submatrix of $X$ without the $j$-th column, and $Re = (Re_1,...,Re_n)^\top $ with $Re_i =  f(u_i)-f(\hat{u}_i) -\dot{f}(\hat{u}_i)(u_i-\hat{u}_i).$ 
We will construct score vector $v_j$ so that the first term on the right hand side of (\ref{decomp1}) is asymptotically normal, while the second and third terms, which together contribute to the bias of the generalized LDP estimator $\check{\beta}_j$, are negligible.

To determine the score vector $v_j$ efficiently, we consider the following node-wise regression among the covariates
\beq \label{eta}
\bx_j = X_{-j}\gamma_j+\eta_j,\quad\quad j=1,...,p,
\eeq
where $\gamma_j = \argmin_{\gamma\in \R^{p-1}}\E[\|\bx_j-X_{-j}\gamma\|_2^2]$ and $\eta_j$ is the error term.
Intuitively, if we set $v_j = \hat{W}^{-1}\eta_j$ for $\hat{W}=\text{diag}({\dot{f}(\hat{u}_1)},...,{\dot{f}(\hat{u}_n)})$, then it should follow that 
\[
\langle v_j,\bold{h}_{-j}\rangle_n \le \max_{k\ne j}| \langle v_j,\bx_k \rangle_n |\cdot\|\hat{\beta}-\beta\|_1 = \max_{k\ne j}| \langle \eta_j,\bx_k \rangle |\cdot\|\hat{\beta}-\beta\|_1 \approx 0. 
\]
In practice, we use the node-wise Lasso to obtain an estimate of $\eta_j$. For $X$ from $\mathcal{D}_2$ and $\hat{\beta}$ obtained from $\mathcal{D}_1$, the score $v_j$ is obtained by calibrating the Lasso-generated residue $\hat{\eta}_j$, i.e.
\[
v_j(\lambda) = \hat{W}^{-1}\hat{\eta}_j(\lambda),\quad \hat{\eta}_j(\lambda) = \bold{x}_j-X_{-j}\hat{\gamma}_j(\lambda),
\]
\beq \label{lasso}
\hat{\gamma}_j(\lambda) = \argmin_b\bigg\{ \frac{\|\bold{x}_j-X_{-j}b\|_2^2}{2n}+\lambda \|b\|_1  \bigg\}.
\eeq
Clearly, $v_j(\lambda)$ depends on the tuning parameter $\lambda$. Define the following quantities
\beq \label{eta,tau}
\zeta_j (\lambda)= \max_{k\ne j}\frac{| \langle v_j(\lambda),\bold{x}_k \rangle_n |}{\|v_j(\lambda)\|_{n}},\quad\quad \tau_j(\lambda) = \frac{\|v_j(\lambda)\|_{n}}{|\langle v_j(\lambda),\bold{x}_j \rangle_n|}.
\eeq
The tuning parameter $\lambda$ can be determined through $\zeta_j (\lambda)$ and $ \tau_j(\lambda)$ by the algorithm in Table 1, which is adapted from the algorithm in \cite{zhang2014confidence}. 

\begin{table}[h!]
	\centering
	\caption{Computation of $v_j$ from the Lasso (\ref{lasso})}
	\begin{tabular*}{ 0.88 \textwidth}{cl}
		\hline 
		Input: &  An upper bound $\zeta_j^*$ for $\zeta_j$, with default value $\zeta^* = \sqrt{2\log p}$, \\
		& tuning parameters $\kappa_0\in[0,1]$ and $\kappa_1\in (0,1]$;\\
		Step 1: & If $\zeta_j(\lambda)>\zeta^*_j$ for all $\lambda>0$, set $\zeta_j^* = (1+\kappa_1)\inf_{\lambda>0}\zeta_j(\lambda)$; \\
		&              $\lambda \leftarrow \max\{\lambda:\zeta_j(\lambda)\le \zeta^*_j\},\zeta^*_j \leftarrow \zeta_j(\lambda),\tau^*_j\leftarrow \tau_j(\lambda)$; \\
		Step 2: &  $\lambda_j \leftarrow \min\{\lambda: \tau_j(\lambda)\le (1+\kappa_0)\tau^*_j\}$;\\
		&               $v_j \leftarrow v_j(\lambda_j), \tau_j \leftarrow \tau_j(\lambda_j),\zeta_j \leftarrow \zeta_j(\lambda_j)$ \\
		Output: & $\lambda_j, v_j, \tau_j,\zeta_j$\\
		\hline
	\end{tabular*}
	\label{table:t1}
\end{table}

Once $\check{\beta}_j$ and $\tau_j$ are obtained, we define the standardized statistics 
\[
M_j = \check{\beta}_j /\tau_j,
\]
for $j=1,...,p$.
The global test statistic is then defined as
\beq \label{test.or}
M_n = \max_{1\le j\le p} M_j^2.
\eeq

\subsection{Asymptotic Null Distribution}

We now turn to the analysis of the properties of the global test statistic $M_n$ defined in \eqref{test.or}.
For the random covariates, we consider both the Gaussian design and the bounded design. 
Under the Gaussian design, the covariates are generated from a multivariate Gaussian distribution with an unknown covariance matrix $\Sig\in \R^{p\times p}$. 
In this case, we assume\\
\indent ({\bf A1}). ${X}_i \sim N(0, \Sig)$ independently for each $i=1,...,n.$  \\
In the case of bounded design, we assume instead \\
\indent ({\bf A2}).  ${X}_i$ for $i=1,...,n$ are i.i.d. random vectors satisfying $\E X_i=0$ and $\max_{1\le i\le n}\|X_i\|_\infty \le T$ for some constant $T>0$. \\
Define the $\ell_1$ ball 
\begin{align*}
\mathcal{B}_1(k) = \bigg\{ \Ome = (\omega_{ij})\in \R^{p\times p}: \max_{1\le i\le p}\sum_{j=1}^p\min\bigg(|\omega_{ij}|\sqrt{\frac{n}{\log p}},1\bigg)\le k \bigg\}.
\end{align*}
In general, $\mathcal{B}_1(k)$ includes any matrix $\Ome$ whose rows $\omega_i$ are $\ell_0$ sparse with $\|\omega_i\|_0\le k$ or $\ell_1$ sparse with $\|\omega_i\|_1\le k\sqrt{{\log p}/{n}}$ for all $i=1, ..., p$. 
The parameter space of the covariance matrix $\Sig$ and the regression vector $\beta$ are defined as following.

({\bf A3}). The parameter space $\Theta(k)$ of $\theta= (\beta, \Sig)\in \R^p\times \R^{p\times p}$ satisfies 
\begin{align*}
\Theta(k) = \bigg\{ (\beta, \Sig): &\quad \|\beta\|_0 \le k, M^{-1}\le \lambda_{\min}(\Sig)\le \lambda_{\max}(\Sig)\le M,  \Sig^{-1}\in \mathcal{B}_1(k) \bigg\},
\end{align*}
for some constant $M\ge 1$. For convenience, we denote $\Theta_1(k)=\{\beta\in\R^p:\|\beta\|_0\le k\}$ and $\Theta_{2}(k)=\{ \Sig\in\R^{p\times p}:M^{-1}\le \lambda_{\min}(\Sig)\le \lambda_{\max}(\Sig)\le M,  \Sig^{-1}\in \mathcal{B}_1(k) \}$, so that $\Theta(k)=\Theta_1(k)\times \Theta_{2}(k)$.


The following theorem states that the asymptotic null distribution of $M_n$ under either the Gaussian or bounded design is a Gumbel distribution.
\bet \label{null.dis}
Let $M_n$ be the test statistic defined in (\ref{test.or}), $D$ be the diagonal of $\Sig^{-1}$ and $(\xi_{ij}) = D^{-1/2}\Sig^{-1}D^{-1/2}$. Suppose $\max_{1\le i<j\le p} |\xi_{ij}|\le c_0$ for some constant $0<c_0<1$, $\log p = O(n^r)$ for some $0<r<1/5$, and
\begin{enumerate}
	\item under the Gaussian design, we assume (A1) (A3) and $k= o\big( \sqrt{n}/\log^3 p\big)$; or
	\item under the bounded design, we assume (A2) (A3) and $k= o\big( \sqrt{n}/\log^{5/2} p\big).$
\end{enumerate}
Then under $H_0$, for any given $x\in \R$, 
\[
P_\theta\big( M_n-2\log p+\log \log p\le x \big) \to \exp\bigg( -\frac{1}{\sqrt{\pi}}\exp (-x/2)\bigg),\quad \text{as $(n,p)\to \infty$}.
\]
\eet

The condition that $\log p = o(n^r)$ for some $0<r<1/5$ is consistent with those required for testing the global hypothesis in high-dimensional linear regression \citep{xia2018two} and for testing two-sample covariance matrices \citep{cai2013two}. It allows the dimension $p$ to be exponentially large comparing to the sample size $n$, which is much more flexible than the likelihood ratio test considered in \cite{sur2017likelihood} and \cite{sur2019modern}, where the dimension can only scale as $p<n$. Under the Gaussian design, it is required that the sparsity $k$ is $o\big( \sqrt{n}/\log^3 p\big)$ whereas for the bounded design, it suffices that the sparsity $k$ to be $o\big( \sqrt{n}/\log^{5/2} p\big)$.

\begin{remark}
	The analysis can be extended to testing $H_0: \beta_G =0$ versus $H_1:\beta_G\ne 0$ for a given index set $G$. Specifically, we can construct the test statistic as $M_{G,n} = \max_{i\in G}M_j^2$ and obtain a similar Gumbel limiting distribution by replacing $p$ by $|G|$, as $(n,|G|)\to \infty$. The sparsity condition thus should be forwarded to the set $G$.
\end{remark}

Based on the limiting null distribution, the asymptotically $\alpha$ level test can be defined as 
\[
\Phi_\alpha(M_n)= I \{ M_n \ge 2\log p-\log\log p+q_\alpha\},
\]
where $q_\alpha$ is the $1-\alpha$ quantile of the Gumbel distribution with the cumulative distribution function $ \exp\big( -\frac{1}{\sqrt{\pi}}\exp (-x/2)\big)$, i.e.
\[
q_\alpha = -\log(\pi)-2\log\log(1-\alpha)^{-1}.
\]
The null hypothesis $H_0$ is rejected if and only if $\Phi_\alpha(M_n)=1$.

\subsection{Minimax Separation Distance and Optimality}

In this subsection, we answer the question: ``What is the essential difficulty for testing the global hypothesis in logistic regression." To fix ideas, we begin with defining the minimax separation distance that measures such an essential difficulty for testing the global null hypothesis at a given level and type II error. In particular, we consider the alternative 
\[
H_1: \beta\in \bigg\{\beta\in \R^p:\|\beta\|_\infty \ge \rho, \|\beta\|_0\le k\bigg\}
\] 
for some $\rho>0$. This alternative concerns the detection of any discernible signals among the regression coefficients where the signals can be extremely sparse, which has interesting applications (see \cite{xia2015testing}). Similar alternatives are also considered by \cite{cai2013two} and \cite{tony2014two}.

By fixing a level $\alpha>0$ and a type II error probability $\delta>0$, we can define the $\delta$-separation distance of a level $\alpha$ test procedure $\Phi_\alpha$ for given design covariance $\Sig$ as
\begin{align} \label{sep.dist}
\rho(\Phi_\alpha,\delta, \Sig) &= \inf \bigg\{\rho>0: \inf_{\beta\in \Theta_1(k):\|\beta\|_\infty \ge \rho} P_{\theta}(\Phi_\alpha = 1)\ge 1-\delta   \bigg\}\nonumber\\
& = \inf \bigg\{\rho>0: \sup_{\beta\in \Theta_1(k):\|\beta\|_\infty \ge \rho} P_{\theta}(\Phi_\alpha = 0)\le \delta   \bigg\}.
\end{align}
The $\delta$-separation distance $\rho(\Phi_\alpha,\delta, \Theta(k))$ over $\Theta(k)$ can thus be defined by taking the supremum over all the covariance matrices $\Sig\in \Theta_{2}(k)$, so that
\begin{align*}
\rho(\Phi_\alpha,\delta, \Theta(k)) &=\sup_{\Sig\in \Theta_{2}(k)} \rho(\Phi_\alpha,\delta, \Sig),
\end{align*}
which corresponds to the minimal $\ell_\infty$ distance such that the null hypothesis $H_0$ is well separated from the alternative $H_1$ by the test $\Phi_\alpha$. In general, $\delta$-separation distance is an analogue of the statistical risk in estimation problems. It characterizes the performance of a specific $\alpha$-level test with a guaranteed type II error $\delta$. Consequently, we can define the $(\alpha,\delta)$-minimax separation distance over $\Theta(k)$ and all the $\alpha$-level tests as
\[
\rho^*(\alpha, \delta, \Theta(k)) = \inf_{\Phi_\alpha}\rho(\Phi_\alpha, \delta, \Theta(k)).
\]
The definition of $(\alpha,\delta)$-minimax separation distance generalizes the ideas of \cite{ingster1993asymptotically}, \cite{baraud2002non} and \cite{verzelen2012minimax}.
The following theorem establishes the minimax lower bound of the $(\alpha,\delta)$-separation distance under the Gaussian design for testing the global null hypothesis over the parameter space $\Theta'(k)\subset \Theta(k)$ defined as
\[
\Theta'(k)= \big(\Theta_1(k)\cap \{ \beta\in \R^p: \|\beta\|_2\lesssim (n^{1/4}\log p)^{-1} \}\big)\times \Theta_2(k).
\]

\bet \label{global.lower}
Assume that $\alpha+\delta\le 1$. Under the Gaussian design, if (A1) and (A3) hold, $(\beta,\Sig)\in \Theta'(k)$ and $k \lesssim \min\{p^\gamma, \sqrt{n}/\log^3 p\}$ for some $0< \gamma <1/2$,
then the $(\alpha,\delta)$-minimax separation distance over $\Theta'(k)$ has the lower bound
\beq \label{global.lower.equation}
\rho^*(\alpha, \delta, \Theta'(k)) \ge c\sqrt{\frac{\log p}{n} }
\eeq
for some constant $c>0$.
\eet
In order to show the above lower bound is asymptotically sharp, we prove that it is actually attainable under certain circumstances, by our proposed global test $\Phi_\alpha$. In particular, for the bounded design, we make the following additional assumption. \\
\indent ({\bf A4}).  It holds that $P_\theta(\max_{1\le i\le n}|\beta^\top X_i| \ge C) =O(p^{-c})$ for some constant $C,c>0$.

\bet \label{power}
Suppose that $\log p = O(n^r)$ for some $0<r<1$. Under the alternative $H_1: \|\beta\|_\infty \ge c_2\sqrt{\log p/{n}}$ for some $c_2>0$, and 
\begin{enumerate}
	\item[(i)] under the Gaussian design, assume that (A1) and (A3) hold, $\|\beta\|_2\le C(\log\log p)/\sqrt{\log n}$ for $C\le \min\{\sqrt{2/\lambda_{\max}(\Sig)}, (2r\sqrt{2\lambda_{\max}(\Sig)})^{-1} \}$, $\log p\gtrsim \log^{1+\delta} n$ for some $\delta>0$ and $k =o({\sqrt{n}/\log^3 p})$; or
	\item[(ii)] under the bounded design, assume  that (A2), (A3), and (A4) hold, and $k =o({\sqrt{n}/\log^{5/2} p})$.
\end{enumerate}
Then we have $P_{\theta}\big( \Phi_\alpha(M_n)=1\big) \to 1$ as  $(n,p)\to \infty$.
\eet

In Theorem \ref{power}, (A4) is assumed for the bounded case and $\|\beta\|_2 = O(\log\log p/\sqrt{\log n})$ is required for the Gaussian case. In particular, since $\log p =O( n^r)$ for some $0<r<1$, the upper bound ${\log\log p}/{\sqrt{\log n}}$ for $\|\beta\|_2$ can be as large as $\sqrt{\log n}$. In Theorem \ref{global.lower}, the minimax lower bound is established over $(\beta,\Sig)\in\Theta'(k)$, so that the same lower bound holds over a larger set
\beq \label{set.pwe}
(\beta,\Sig)\in \big(\Theta_1(k)\cap \{ \beta\in \R^p: \|\beta\|_2\le \log\log p/\sqrt{\log n} \}\big)\times \Theta_2(k),
\eeq
since $\log\log p/\sqrt{\log n}\gtrsim  (n^{1/4}\log p)^{-1}$. On the other hand, Theorem \ref{power} (i) indicates an upper bound $\rho^*\lesssim \sqrt{\log p/n}$ attained by our proposed test under the Gaussian design over the set (\ref{set.pwe}). These two results  imply the minimax rate $\rho^*\asymp \sqrt{\log p/n}$ and the minimax optimality of our proposed test over the set (\ref{set.pwe}).

\subsection{Comparison with  Existing Works}

In this section, we make detailed comparisons and connections with some existing works concerning global hypothesis testing in the high-dimensional regression literature.

\cite{ingster2010detection} addressed the detection boundary for high-dimensional sparse linear regression models, and more recently \cite{mukherjee2015hypothesis} studied the detection boundary for hypothesis testing in high-dimensional sparse binary regression models. However, although both works obtained the sharp detection boundary for the global testing problem $H_0: \beta=0$, their alternative hypotheses are different from ours. Specifically,  \cite{mukherjee2015hypothesis} considered the alternative hypothesis $H_1: \beta\in\big\{ \beta\in \R^p: \|\beta\|_0\ge k, \min\{|\beta_j|:\beta_k\ne 0\}\ge A \big\}$, which implies that 
$\beta$ has at least $k$ nonzero coefficients exceeding $A$ in absolute values.  \cite{ingster2010detection} considered the alternative hypothesis $
H_1: \beta\in \big\{ \beta\in \R^p: \|\beta\|_0\le k, \|\beta\|_2\ge \rho \big\}$, which 
concerns $k$ sparse $\beta$ with $\ell_2$ norm at least $\rho$. In fact, the proof of our Theorem 2 can be directly extended to such an alternative concerning the $\ell_2$ norm, which amounts to obtaining a lower bound of order $\sqrt{\frac{k\log p}{n}}$ for high dimensional logistic regression. However, developing a minimax optimal test for such alternative is beyond the scope of the current paper. 

Additionally,  in contrast to the minimax separation distance considered in this paper,  the papers  by  \cite{ingster2010detection} and \cite{mukherjee2015hypothesis} considered the minimax risk (or the minimax total error probability)  given by
\beq
\inf_{\Phi}\sup_{\Sig\in\Theta_{2}(k)}\text{Risk}(\Phi,\Sig)=\inf_{\Phi}\sup_{\Sig\in\Theta_{2}(k)}\bigg\{\max_{\beta\in H_0}P_{\theta}(\Phi=1)+\max_{\beta\in \Theta_1(k):\|\beta\|_\infty\ge \rho}P_\theta(\Phi=0)\bigg\},
\eeq
where the infimum is taken over all tests $\Phi$. This  minimax risk can be also written as
\beq \label{mini.risk}
\inf_{\Phi}\sup_{\Sig\in\Theta_{2}(k)}\text{Risk}(\Phi,\Sig)=\inf_{\alpha\in(0,1)}\bigg\{\alpha+\inf_{\Phi_\alpha}\sup_{\Sig\in\Theta_{2}(k)}\sup_{\beta\in \Theta_{1}(k):\|\beta\|_\infty\ge \rho}P_\theta(\Phi_\alpha=0)\bigg\}.
\eeq
A comparison of (\ref{sep.dist}) and (\ref{mini.risk}) yields the slight difference between the two criteria, as one depends on a given Type I error $\alpha$ and the other doesn't.

Moreover, these two papers considered different design scenarios from ours. In \cite{ingster2010detection}, only the isotropic Gaussian design was considered. As a result, the optimal tests proposed therein rely highly on the independence assumption. In \cite{mukherjee2015hypothesis}, the general binary regression was studied under fixed sparse design matrices. In particular, the minimax lower and upper bounds were  only derived in the special case of design matrices with binary entries and certain sparsity structures. 

In comparison with the recent works of \cite{sur2017likelihood}, \cite{candes2018phase} and \cite{sur2019modern}, besides the aforementioned difference in the asymptotics of $(p,n)$,  these two papers only considered  the random Gaussian design, whereas our work also considered random bounded design as in \cite{van2014asymptotically}.  In addition,   \cite{sur2017likelihood} and \cite{sur2019modern} developed  the Likelihood Ratio (LLR) Test for testing the hypothesis $H_0: \beta_{j_1}=\beta_{j_2}=...=\beta_{j_k}=0$ for any finite $k$. Intuitively, a valid test for the global null and $p/n\to \kappa\in(0,1/2)$ can be adapted from the individual LLR tests using the Bonferroni procedure. However, as our simulations show (Section 5), such a test is less powerful compared  to our proposed test.

Lastly, our  minimax  results focus on the highly sparse regime $k \lesssim p^{\gamma}$ where $\gamma\in(0,1/2)$. As shown by \cite{ingster2010detection} and \cite{mukherjee2015hypothesis}, the problem under the dense regime where $\gamma\in(1/2,1)$ can be very different from the sparse regime. Mostly likely, the fundamental difficulty of the testing problem changes in this situation so that different methods need to be carefully developed. We leave these interesting questions for future investigations.

\section{LARGE-SCALE MULTIPLE TESTING}

Denote by $\beta$ the true coefficient vector in the model and denote $\HH_0 = \{j: \beta_j=0, j=1,\cdots, p\}, \HH_1 = \{j:\beta_j\ne 0, j=1,\cdots, p\}$. In order to identify the indices in $\HH_1$, we consider simultaneous testing of the following null hypotheses
\[
H_{0,j}: \beta_j=0 \quad\text{vs.} \quad H_{1,j}:\beta_j \ne 0, \quad1\le j\le p. 
\]
Apart from identifying as many nonzero $\beta_j$ as possible, to obtain results of practical interest, we would like to control the false discovery rate (FDR) as well as the false discovery proportion (FDP), or the number of falsely discovered variables (FDV).

\subsection{Construction of Multiple Testing Procedures}

Recall that in Section 2, we define the standardized statistics $M_j = \check{\beta}_j /\tau_j,$
for $j=1,...,p$. For a given threshold level $t>0$, each individual hypothesis $H_{0,j}: \beta_j=0$ is rejected if $|M_j|\ge t$. Therefore for each $t$, we can define 
\[
\text{FDP}_\theta(t) = \frac{\sum_{j\in \mathcal{H}_0} I\{ |M_{j}| \ge t\}}{\max \big\{\sum_{j=1}^p I\{ |M_{j}| \ge t\},1   \big\}}, \quad\quad \text{FDR}_\theta(t) = \E_\theta [\text{FDP}(t)],
\]
and the expected number of falsely discovered variables $\text{FDV}_\theta(t) = \E_\theta \big[\sum_{j\in \mathcal{H}_0} I\{ |M_{j}| \ge t\}\big].$
\paragraph{Procedure Controlling FDR/FDP.} In order to control the FDR/FDP at a pre-specified level $0<\alpha <1$, we can set the threshold level as
\beq
\tilde{t}_1 = \inf\bigg\{ 0\le t\le b_p: \frac{\sum_{j\in \mathcal{H}_0} I\{ |M_{j}| \ge t\}}{\max \big\{\sum_{j=1}^p I\{ |M_{j}| \ge t\},1   \big\}} \le \alpha  \bigg\},
\eeq
for some $b_p$ to be determined later.

In general, the ideal choice $\tilde{t}_1$ is unknown and needs to be estimated because it depends on the knowledge of the true null $\mathcal{H}_0$. Let $G_0(t)$ be the proportion of the nulls falsely rejected by the procedure among all the true nulls at the threshold level $t$, namely, $G_0(t) = \frac{1}{p_0}\sum_{j\in \mathcal{H}_0}I\{ |M_{j}| \ge t\}, $
where $p_0 = |\mathcal{H}_0|$. In practice, it is reasonable to assume that the true alternatives are sparse. If the sample size is large, we can use the tails of normal distribution $G(t) = 2-2\Phi(t)$ to approximate $G_0(t)$. In fact, it will be shown that, for $b_p=\sqrt{2\log p-2\log \log p}$, $\sup_{0\le t\le b_{p}} \big| \frac{G_0(t)}{G(t)}-1 \big| \rightarrow 0$ in probability as $(n,p) \rightarrow \infty$. To summarize, we have the following logistic multiple testing (LMT) procedure controlling the FDR and the FDP.

\begin{proc}[LMT]
	Let $0<\alpha <1$, $b_p=\sqrt{2\log p-2\log\log p}$ and define
	\beq \label{t.hat.fdr}
	\hat{t} = \inf\bigg\{ 0\le t\le b_p: \frac{pG(t)}{\max \big\{\sum_{j=1}^p I\{ |M_{j}| \ge t\},1   \big\}} \le \alpha  \bigg\}.
	\eeq
	If $\hat{t}$ in (\ref{t.hat.fdr}) does not exist, then let $\hat{t} = \sqrt{2\log p}$. We reject $H_{0,j}$ whenever $|M_{j}| \ge \hat{t}$.
\end{proc}

\paragraph{Procedure Controlling FDV.} For large-scale inference, it is sometimes of interest to directly control the number of falsely discovered variables (FDV) instead of the less stringent FDR/FDP, especially when the sample size is small \citep{liu2014hypothesis}.  By definition, the FDV control, or equivalently, the per-family error rate control, provides an intuitive description of the Type I error (false positives) in variable selection. Moreover, controlling $\text{FDV}=r$ for some $0<r<1$ is related to the family-wise error rate (FWER) control, which is the probability of at least one false positive. In fact, FDV control can be achieved by a suitable modification of the FDP controlling procedure introduced above.
Specifically, we propose the following FDV (or FWER) controlling logistic multiple testing (LMT$_V$) procedure.

\begin{proc}[LMT$_V$]
	For a given tolerable number of falsely discovered variables $r < p$ (or a desired level of FWER $0<r<1$), let $\hat{t}_{FDV} = G^{-1}(r/p).$
	$H_{0,j}$ is rejected whenever $|M_{j}| \ge \hat{t}_{FDV}$.
\end{proc}

\subsection{Theoretical Properties for Multiple Testing Procedures}

In this section we show that our proposed multiple testing procedures control the theoretical FDR/FDP or FDV asymptotically. For simplicity, our theoretical results are obtained under the bounded design scenario. For FDR/FDP control, we need  an  additional assumption on the interplay between the dimension $p$ and the parameter space $\Theta(k)$.

Recall that $\eta_j=(\eta_{j1},...,\eta_{jn})^\top$ for $j=1,...,p$ defined in (\ref{eta}). We define $F_{jk} = \E_\theta [\eta_{ij}\eta_{ik}/\dot{f}(u_i)]$ for $1\le j,k\le p$, and $\rho_{jk} = F_{jk}/\sqrt{F_{jj}F_{kk}}$. Denote $\mathcal{B}(\delta) = \{ (j,k): |\rho_{jk}| \ge \delta, i \ne j  \}$ and $\mathcal{A}(\epsilon) = \mathcal{B}((\log p)^{-2-\epsilon}).$\\
\indent ({\bf A5}). Suppose that for some $\epsilon >0$ and $q>0$, $\sum_{(j,k)\in \mathcal{A}(\epsilon): j,k\in \HH_0} p^{\frac{2|\rho_{jk}|}{1+|\rho_{jk}|}+q} = O(p^2/(\log p)^2).$

The following proposition shows that $M_{j}$ is asymptotically normal distributed and $G_0(t)$ is well approximated by $G(t)$. 
\begin{proposition} \label{prop.1}
	Under (A2) (A3) and (A4), suppose $p =O( n^c)$ for some constant $c>0$, $k=o(\sqrt{n}/\log^{5/2} p)$, then as $(n,p)\rightarrow \infty$, 
	\beq \label{prop1}
	\sup_{j\in \mathcal{H}_0} \sup_{0\le t\le \sqrt{2\log p}} \bigg| \frac{P_\theta(|M_{j}|\ge t)}{2-2\Phi(t)}-1  \bigg| \rightarrow 0.
	\eeq
	If in addition we assume (A5), then
	\beq \label{prop2}
	\sup_{0\le t\le b_{p}} \bigg| \frac{G_0(t)}{G(t)}-1 \bigg| \rightarrow 0
	\eeq
	in probability, where $\Phi$ is the cumulative distribution function of the standard normal distribution and $b_p=\sqrt{2\log p-2\log\log p}$.
\end{proposition}

The following theorem provides the asymptotic FDR and FDP control of our procedure.
\bet \label{fdr}
Under the conditions of Proposition \ref{prop.1}, for $\hat{t}$ defined in our LMT procedure, we have
\beq
\lim_{(n,p)\rightarrow \infty} \frac{\textup{FDR}_\theta(\hat{t})}{\alpha p_0/p} \le 1, \quad\quad \lim_{(n,p)\rightarrow \infty} P_\theta\bigg(\frac{\textup{FDP}_\theta(\hat{t})}{\alpha p_0/p}\le 1+\epsilon \bigg) = 1
\eeq
for any $\epsilon>0$. 
\eet

For the FDV/FWER controlling procedure, we have the following theorem.
\bet \label{fdv}
Under (A2) (A3) and (A4), assume $p=O(n^c)$ for some $c>0$ and $k=o(\sqrt{n}/\log^{5/2} p)$. Let $r<p$ be the desired level of FDV. For $\hat{t}_{FDV}$ defined in our LMT$_V$ procedure, we have $\lim_{(n,p)\rightarrow \infty} \frac{\textup{FDV}_\theta(\hat{t}_{FDV})}{rp_0/p} \le 1.$
In addition, if $0<r<1$, we have $
\lim_{(n,p)\rightarrow \infty}\frac{\textup{FWER}_\theta(\hat{t}_{FDV})}{rp_0/p}\le 1.$
\eet

The above theoretical results are obtained under the dimensionality condition  $p=O(n^c)$, which is stronger than that of the global test. Essentially, the condition is needed to obtain the uniform convergence (\ref{prop1}), whose form (as ratio) is stronger than the convergence in distribution in the ordinary sense (as direct difference).

\section{TESTING FOR TWO LOGISTIC REGRESSION MODELS}

In some applications, it is also interesting to consider hypothesis testing that involves two separate logistic regression models of the same dimension.   Specifically, for $\ell=1,2$ and $i=1,...,n_{\ell}$, where $n_1\asymp n_2$, $y^{(\ell)}_i = f({\beta^{{(\ell)}}}^\top X_i^{(\ell)})+\epsilon_i^{(\ell)}$,
where $f(u)={e^u}/({1+e^u})$, and $\epsilon_i^{(\ell)}$ is a binary random variable such that $y_i^{(\ell)}| X_i^{(\ell)} \sim \text{Bernoulli}(f({\beta^{{(\ell)}}}^\top X_i^{(\ell)}))$.
The global   null hypothesis $H_0: \beta^{(1)}=\beta^{(2)}$ implies that there is overall no difference in association between covariates and the response. If this null hypothesis is rejected,  we are interested in simultaneously testing the hypotheses $H_{0,j}: \beta_j^{(1)} =\beta_j^{(2)}$ for each $j=1,...,p$. 

To test  the global null $H_0: \beta^{(1)}=\beta^{(2)}$ against $H_1: \beta^{(1)}\ne\beta^{(2)},$
we can first obtain $\check{\beta}_j^{(\ell)}$ and $\tau_j^{(\ell)}$ for each model, and then calculate the coordinate-wise standardized statistics $T_j = \frac{\check{\beta}_j^{(1)}}{\sqrt{2}\tau_j^{(1)}} -\frac{\check{\beta}_j^{(2)}}{\sqrt{2}\tau_j^{(2)}},$ for $j=1,...,p$.
Define 
the global test statistic as
$
T_n = \max_{1\le j\le p} T_j^2,$ it can be shown that the limiting null distribution is also a Gumbel distribution. The $\alpha$ level global test is thus defined as $\Phi_\alpha(T_n)= I \{ T_n \ge 2\log p-\log\log p+q_\alpha\}$, where $q_\alpha = -\log(\pi)-2\log\log(1-\alpha)^{-1}.$ 
For  multiple  hypotheses testing of two regression vectors $H_{0,j}: \beta_j^{(1)} =\beta_j^{(2)}$ for $j=1,...,p,$
we  consider the test statistics $T_j$ defined above. The two-sample multiple testing procedure controlling FDR/FDP is given as follows. 

\begin{proc}
	Let $0<\alpha <1$ and define $\hat{t} = \inf\bigg\{ 0\le t\le b_p: \frac{pG(t)}{\max \big\{\sum_{j=1}^p I\{ |T_{j}| \ge t\},1   \big\}} \le \alpha  \bigg\}.$
	If the above $\hat{t}$ does not exist,  let $\hat{t} = \sqrt{2\log p}$. We reject $H_{0,j}$ whenever $|T_{j}| \ge \hat{t}$. 
\end{proc}

\section{SIMULATION STUDIES}

In this section we examine the numerical performance of the proposed tests. Due to the space limit, for both global and multiple testing problems, we only focus on the single regression setting, and report the results on two logistic regressions in the  Supplementary Materials. Throughout our numerical studies, sample splitting was not used.

\subsection{Global Hypothesis Testing}

In the following simulations, we consider a variety of dimensions, sample sizes, and sparsity of the models. For alternative hypotheses, the dimension of the covariates $p$ ranges from 100, 200, 300 to 400, and the sparsity $k$ is set as 2 or 4. The sample sizes $n$ are determined by the ratio $r=p/n$ that takes values of 0.2, 0.4 and 1.2. To generate the design matrix $X$, we consider the Gaussian design with the blockwise-correlated covariates so that $\Sig=\Sig_B$, where $\Sig_B$ is a $p\times p$ blockwise diagonal matrix including 10 equal-sized blocks, whose diagonal elements are $1$'s and off-diagonal elements are set as $0.7$. Under the alternative, suppose $\mathcal{S}$ is the support of the regression coefficients $\beta$ and $|\mathcal{S}|=k$, we set $|\beta_j| = \rho1\{ j\in \mathcal{S}\}$ for $j=1,...,p$ and $\rho=0.75$ with equal proportions of $\rho$ and $-\rho$.  We set $\kappa_0 = 0$ and $\kappa_1=0.5$. 

To assess the empirical performance of our proposed test ("Proposed"), we compare our test with (i) a Bonferroni procedure applied to the p-values from univariate screening using MLE statistic ("U-S"), and (ii) to the method of \cite{sur2017likelihood,sur2019modern} ("LLR") in the setting where $r=0.2$ and 0.4.

\begin{figure}[h!]
	\centering
	\includegraphics[angle=0,width=16cm]{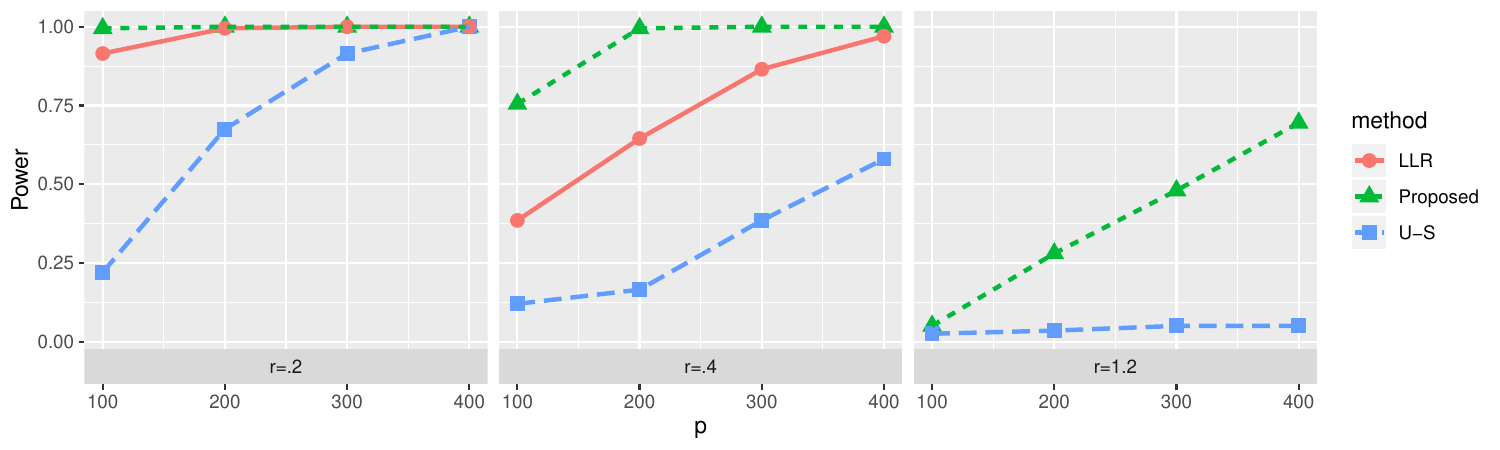}
	\includegraphics[angle=0,width=16cm]{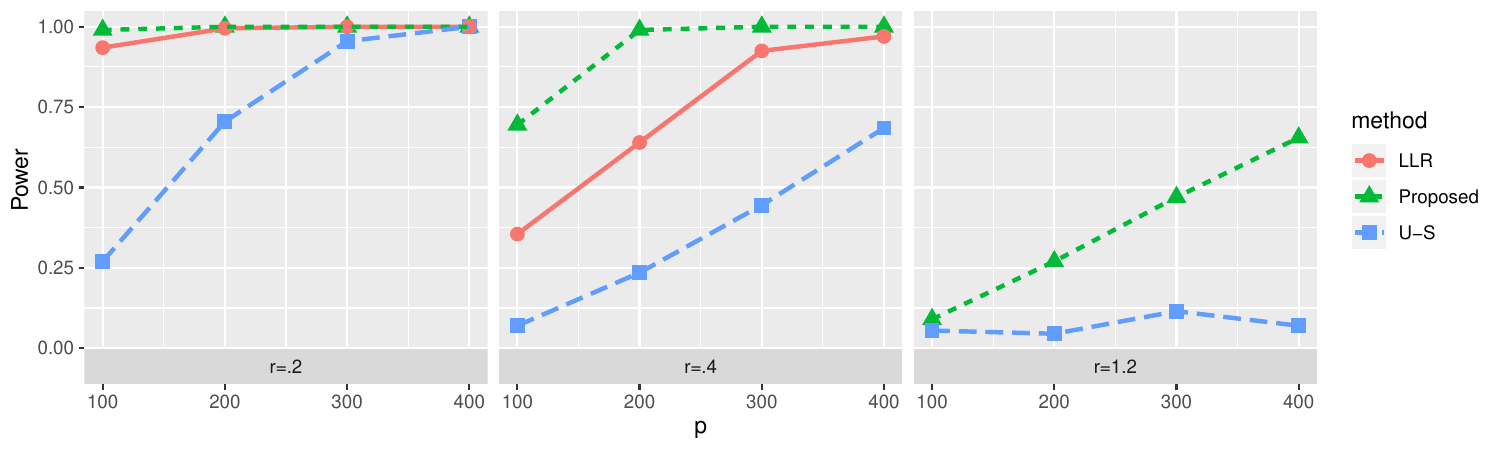}
	\caption{Empirical power with $\alpha=0.05$ for the proposed method (Proposed), the Bonferroni corrected univariate screening method (U-S) and the Bonferroni corrected likelihood ratio based method of \cite{sur2019modern} (LLR). Top panel: $k=2$; bottom panel: $k=4$. } \label{one.pow}
\end{figure}

Table \ref{table:t0} shows the empirical type I errors of these tests at level $\alpha =0.05$ based on 1000 simulations. 
Figure \ref{one.pow} shows the corresponding empirical powers under various settings.  As we expected, our proposed method outperforms the other two alternatives in all the cases (including the moderate dimensional cases where $r=0.2$ and 0.4), and the power increases as $n$ or $p$ grows. In the rather lower dimensional setting where $r=0.2$, the LLR performs almost as well as our proposed method.

\begin{table}[h!]
	\centering
	\caption{Type I error with $\alpha=0.05$ for the proposed method (Proposed), the Bonferroni corrected univariate screening method (U-S) and the Bonferroni corrected likelihood ratio based method of \cite{sur2019modern} (LLR), for different $n$ and $p$.}
	\begin{tabular}{cccccccc}
	\hline
	$p/n$ 	&   $p=100$  & 200 & 300 & 400  & 600 & 800 & 1000  \\  
	\hline  
	\multicolumn{8}{c}{Proposed}\\
	0.2 &    0.052  & 0.066 &0.042 & 0.054 &0.050&0.046&0.070  \\
	0.4 & 0.038 & 0.054 & 0.062 & 0.054   &0.050&0.060&0.074   \\
	1.2 &  0.026& 0.044 & 0.042  & 0.045  &0.044&0.054&0.054 \\
	\multicolumn{8}{c}{U-S}\\
	0.2 &    0.040  & 0.032 &0.024 & 0.018 &0.022&0.028&0.034  \\
	0.4 & 0.050 & 0.032 & 0.024 & 0.020  &0.028&0.032&0.046   \\
	1.2 &  0.028& 0.038 & 0.024  & 0.020    &0.018&0.034&0.014 \\
	\multicolumn{8}{c}{LLR}\\
	0.2 &    0.050  & 0.050 &0.068 & 0.040&0.044&0.046&0.034  \\
	0.4 & 0.084 & 0.070 & 0.048 & 0.056 &0.042&0.058&0.064   \\
	\hline
\end{tabular}
	\label{table:t0}
\end{table}

\subsection{Multiple Hypotheses Testing}

\paragraph{FDR Control.}  In this case, we set $p=800$ and let $n$ vary from 600, 800, 1000, 1200 to 1400, so that all the cases are high-dimensional in the sense that $p>n/2$. The sparsity level $k$ varies from 40, 50 to 60. For the true positives, given the support $\mathcal{S}$ such that $|\mathcal{S}|=k$, we set $|\beta_j| =\rho1\{ j\in \mathcal{S}\}$ for $j=1,...,p$ with equal proportions of $\rho$ and $-\rho$. The design covariates $X_i$'s are generated from a $(|X_i^\top \beta|<3)$-truncated multivariate Gaussian distribution with covariance matrix $\Sig=0.01\Sig_M$, where $\Sig_M$ is a $p\times p$ blockwise diagonal matrix of $10$ identical  unit diagonal Toeplitz matrices whose off-diagonal entries descend from 0.1 to 0 (see Supplementary Material for the explicit form). The choice of  $\kappa_0$ and $\kappa_1$ are the same as the global testing. Throughout, we set the desired FDR level as $\alpha=0.2$.

\begin{figure}[h!]
	\centering
	\includegraphics[angle=0,width=10cm]{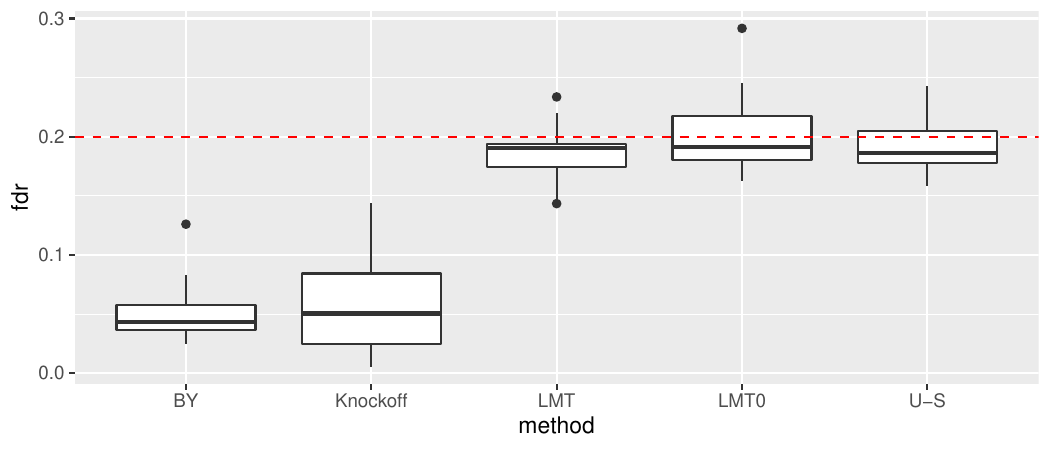}
	\caption{Boxplots of the empirical FDRs across all the settings for $\alpha=0.2$.}\label{fdr.box}
\end{figure} 

\begin{figure}[h!]
	\centering
	\includegraphics[angle=0,width=16cm]{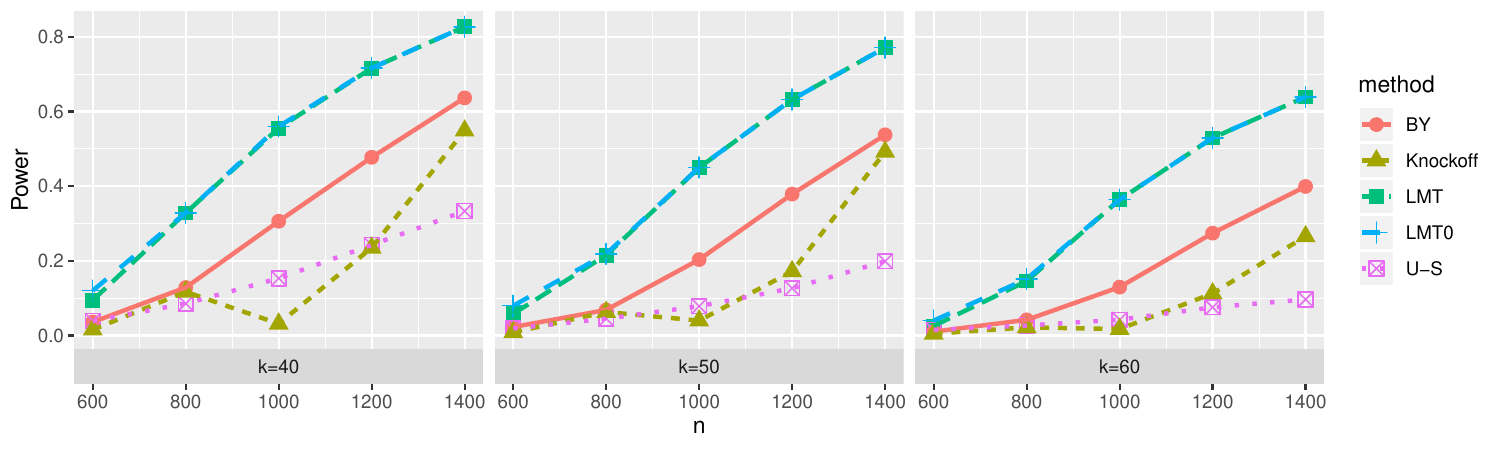}
	\includegraphics[angle=0,width=16cm]{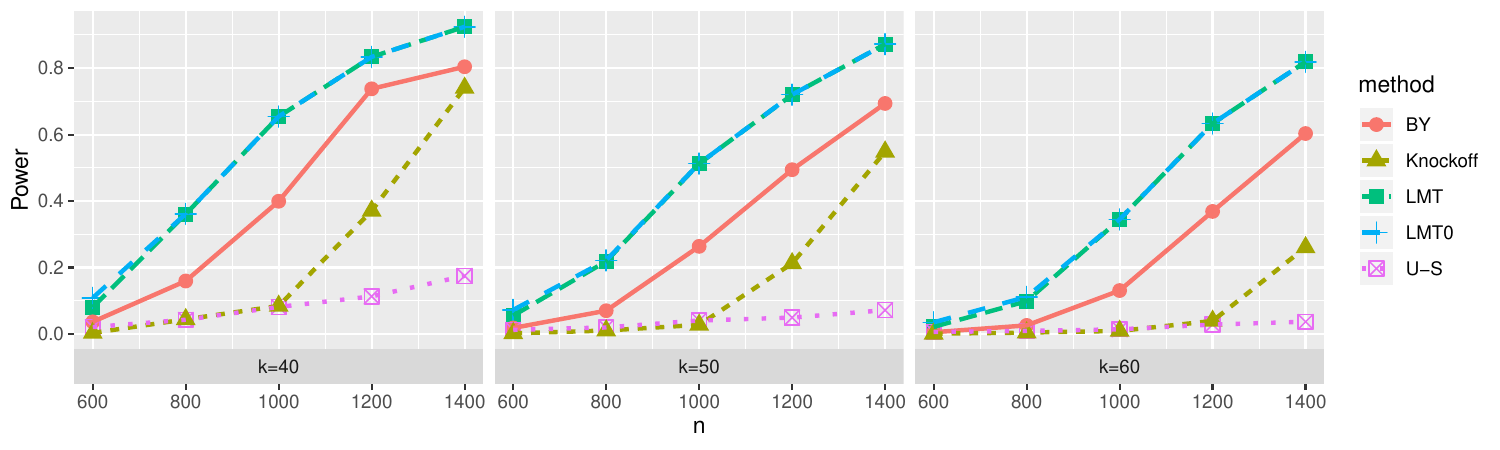}
	\caption{Empirical power under FDR $\alpha=0.2$ for $\rho=3$ (top) and $\rho=4$ (bottom).}\label{pow.res}
\end{figure} 

We compare our proposed procedure (denoted as "LMT") with following methods: (i) the basic LMT procedure with $b_p$ in (\ref{t.hat.fdr}) replaced by $\infty$  ("LMT0"), which is equivalent to applying the BH procedure \citep{benjamini1995controlling} to our debiased statistics $M_j$, (ii) the BY procedure \citep{benjamini2001control} using our debiased statistics $M_j$ ("BY"), implemented using the \texttt{R} function \texttt{p.adjust(...,method="BY")}, (iii) a BH procedure applied to the p-values from univariate screening using the MLE statistics ("U-S"), and (iv) the knockoff method of \cite{candes2018panning} ("Knockoff"). Figure \ref{fdr.box} shows boxplots of the pooled empirical FDRs (see Supplementary Material for the case-by-case FDRs) and Figure \ref{pow.res} shows the empirical powers of these methods based on 1000 replications. Here the power is defined as the number of correctly discovered variables divided by the number of truly associated variables. As a result, we find that LMT and LMT0 correctly control FDRs and have the greatest power among all the cases. In particular, the power of LMT and LMT0 are almost the same, which increases as the sparsity decreases, the signal magnitude $\rho$ increases, or the sample size $n$ increases, while LMT0 has slightly inflated FDRs. The U-S method, although correctly controls the FDRs, has poor power, which is largely due to the dependence among the covariates.

\paragraph{FDV Control.} For our proposed test that controls FDV (denoted as LMT$_V$), by setting desired FDV level $r=10$, we apply our method to various settings. Specifically, we set $\rho=3$, $p\in\{800, 1000, 1200\}$, set $k\in \{40,50,60\}$, and let $n$ vary from 400, 600, 800 to 1000. The design covariates are generated similarly as the previous part. The resulting empirical FDV and powers are summarized in Table 3. Our proposed LMT$_V$ has the correct control of FDV in all the settings and the power increases as $n$ grows, $k$ decreases, or $p$ decreases.

\begin{table}[h!]
	\centering
	\caption{Empirical performance of LMT$_{V}$ with FDV level $r=10$.}
	\begin{tabular}{ccl|cccc|cccc}
		\hline
		\multirow{2}{1em}{  $\rho$}&\multirow{2}{1em}{$p$}&\multirow{2}{2em}{$k$}&\multicolumn{4}{c}{Empirical FDV} &\multicolumn{4}{c}{Empirical Power}  \\
		\cline{4-11}
		& &&   $n=400$  & 600 & 800 & 1000 & 400  & 600 & 800 & 1000  \\  
		\hline
		& &40 &    4.07  & 5.45 &  6.44 & 7.11 & 0.08& 0.23 & 0.40& 0.59\\
		&800 &50 &  4.30& 6.29& 7.27& 8.26   & 0.06 & 0.16& 0.32& 0.49\\
		& & 60 &  4.33 & 6.63& 7.48& 8.42   & 0.05& 0.12& 0.25 & 0.42\\
		\cline{2-11}
		& &40 &    3.30 &4.59 &5.79& 6.82 &  0.06 & 0.18& 0.35& 0.52\\
		$3$&1000 &50 &3.49 & 5.42 &6.43 &7.03   &  0.05& 0.13& 0.26 &0.43 \\
		& & 60 & 3.68 & 5.47& 7.29& 7.97   &  0.03 & 0.09& 0.20& 0.34\\
		\cline{2-11}
		& &40 &    2.69 & 4.36 &5.00& 5.68 & 0.05 & 0.15& 0.31& 0.46  \\
		&1200 &50 & 2.97 & 4.22& 5.73& 6.43   &0.03 &0.11& 0.21& 0.36 \\
		& & 60 &  2.78 &4.91& 5.91& 7.25   &0.02 & 0.07 &0.16& 0.27\\
		\hline
	\end{tabular}
	\label{table:t2}
\end{table}

\section{REAL DATA ANALYSIS}

We illustrate our proposed methods by analyzing a dataset from the Pediatric Longitudinal Study of Elemental Diet and Stool Microbiome Composition (PLEASE) study, a prospective cohort study  to investigate the effects of inflammation, antibiotics, and diet as environmental stressors on the gut microbiome in pediatric Crohn's disease \citep{lewis2015inflammation,lee2015comparative,ni2017role}. The study considered the  association between pediatric Crohn's disease and fecal metabolomics by collecting  fecal samples of 90 pediatric patients with Crohn's disease at baseline, 1 week, and 8 weeks after initiation of either anti-tumor necrosis factor (TNF) or enteral diet therapy, as well as those from 25 healthy control children \citep{lewis2015inflammation}. In details, an untargeted fecal metabolomic analysis was performed on these samples using liquid chromatography-mass spectrometry (LC-MS). Metabolites with more than 80\% missing values across all samples were removed from the analysis. For each metabolite, samples with the missing values were imputed with its minimum abundance across samples. To avoid potential large outliers, for each sample, the metabolite abundances were further normalized by dividing 90\% cumulative sum of the abundances of all metabolites. The normalized abundances were then log transformed and used in all analyses. The metabololomics  annotation was obtained from Human Metabolome Database \citep{lee2015comparative}. In total, for each sample, abundances of 335 known metabolites were obtained and used in our analysis.

\subsection{Association Between Metabolites and Crohn's Disease Before and After Treatment}
We first test the overall association between 335 characterized metabolites and Crohn's disease by fitting a logistic regression using the  data of  25 healthy controls and 90 Crohn's disease patients at the baseline. We obtain a global test statistic of 433.88 with a p-value $<0.001$, indicating a strong association between Crohn's disease and fecal metabolites.  At the FDR $<5\%$,  our multiple testing procedure selects four metabolites, including 	C14:0.sphingomyelin,  C24:1.Ceramide.(d18:1) and   3-methyladipate/pimelate (see Table 4).  Recent studies have demonstrated that sphingolipid metabolites, particularly ceramide and sphingosine-1-phosphate, are signaling molecules that regulate a diverse range of cellular processes that are important in immunity, inflammation and inflammatory disorders \citep{maceyka2014sphingolipid}. In fact, ceramide acts to reduce tumor necrosis factor (TNF) release \citep{rozenova2010studies} and has important roles in the control of autophagy, a process strongly implicated in the pathogenesis of Crohn's disease \citep{barrett2008genome,sewell2012lipidomic}.

We next investigate whether treatment of Crohn's disease alters the association between metabolites and Crohn's disease by fitting two separate logistic regressions using the metabolites measured one week or 8 weeks after the treatment. At each time point, a significant association is detected based on our global test (
p-value $<0.001$).  One week after the treatment, we observe six metabolites associated with Crohn's disease, including all four identified at the baseline and two additional metabolites, beta-alanine and 
adipate (see Table 4). The beta-alanine and adipate associations are likely due to that beta-alanine and adipate  are  important ingredients of the enteral nutrition  treatment of Crohn's disease.  However, it is interesting that at 8 weeks after the  treatment, valine, 
C16.carnitine and  C18.carnitine  are identified to be associated with Crohn's disease together with  3-methyladipate/pimelate and  beta-alanine.  It is known that carnitine plays an important role in Crohn's disease, which might be a consequence of the underlying functional association between Crohn's disease and mutations in the carnitine transporter genes \citep{peltekova2004functional, fortin2011carnitine}. Deficiency of carnitine can lead to severe gut atrophy, ulceration and inflammation in animal models of carnitine deficiency  \citep{carnitine0}. Our results may suggest that the treatment increases carnitine, leading to reduction of inflammation.

\begin{table}[h!]
	\centering
	\caption{Significant metabolites associated with Crohn's disease (coded as 1 in logistic regression) at the baseline, one week and 8 weeks after treatment with FDR $<5\%$. The refitted regression coefficients show the direction of the  association.  }
	\begin{tabular}{cclc}
		\hline
		Disease Stage &HMDB ID & Synonyms & Refitted Coefficient \\
		\hline
		\multirow{4}{4em}{Baseline}&  00885 & C16:0.cholesteryl ester   & 4.45  \\
		& 12097 &C14:0.sphingomyelin &  1.74 \\
		& 04953 & C24:1.Ceramide.(d18:1)&  4.25  \\     
		& 00555 & 3-methyladipate/pimelate&  -12.82 \\
		\hline
		\multirow{6}{4em}{Week 1}& 06726 & C20:4.cholesteryl ester  & 2.17\\
		& 12097 &C14:0.sphingomyelin & 2.06 \\
		& 04949 & C16:0.Ceramide.(d18:1) & 0.87\\
		& 00555 & 3-methyladipate/pimelate & -6.10\\          
		& 00056 & beta-alanine & 2.95\\
		& 00448 & adipate & -4.50 \\
		\hline
		\multirow{5}{4em}{Week 8}& 00883 & valine & 1.40 \\
		& 00222 & C16.carnitine & 0.58 \\
		& 00848 & C18.carnitine & 0.39 \\    
		& 00555 & 3-methyladipate/pimelate & -5.95\\    
		& 00056 & beta-alanine & 0.63\\
		\hline
	\end{tabular}
	\label{multiple.real.1}
\end{table}

\subsection{Comparison of Metabolite Associations Between Responders and Non-Responders}

To compare the metabolic association with Crohn's disease for responders  ($n=47$) and   non-responders  ($n=34$)   eight weeks after treatment, we fit  two logistic regression models, responder versus normal control  and non-responder versus normal control. Our global test shows that there is an overall difference in regression coefficients for responders and for non-responders when compared to the normal controls (p-value $ < 0.001$). We next apply our proposed multiple testing procedure to identify the metabolites that have different regression coefficients in these two different logistic regression models. At the FDR $<0.05$, our procedure identifies 9 metabolites with different regression coefficients (see Table \ref{multiple.real.2}). It is interesting that all these 9 metabolites have the same signs of the refitted coefficients, while the actual magnitudes of the associations between responders and non-responders when compared to the normal controls are different. 
Besides C24:4.cholesteryl ester,
beta-alanine,
valine,
C18.carnitine  and  3-methyladipate/pimelate that we observe in previous analyses,  metabolites 5-hydroxytryptopha,
nicotinate,  and succinate also have differential associations between responders and non-responders when compared to the controls. 

\begin{table}[h!]
	\centering
	\caption{Significant metabolites identified via logistic  regression of responder vs normal control and non-responder vs normal control  for FDR $\le 5\%$.}
	\begin{tabular}{clcc}
		\hline
		\multirow{2}{5em}{HMDB ID}&\multirow{2}{4.5em}{Synonyms} &  \multicolumn{2}{c}{Refitted Coefficients} \\
		\cline{3-4}
		&&  Responder vs.& Non-Responder vs. \\
		&&  Normal &  Normal  \\
		\hline
		06726 & C20:4.cholesteryl ester & 0.139 & 1.854 \\
		01043 & Linoleic.acid &-0.686 & -0.388 \\  
		00472 & 5-hydroxytryptophan & 1.000 & 1.034 \\        
		00056 &  beta-alanine &0.503 &   2.298    \\               
		00883 & valine  &0.628 &   0.530   \\                     
		00848 & C18.carnitine  & 1.100 & 0.457 \\  
		01488 & nicotinate & -1.936 &   -4.312  \\             
		00254 & succinate & 0.750  &  1.508   \\              
		00555 & 3-methyladipate/pimelate &-1.989 & -4.209 \\
		\hline
	\end{tabular}
	\label{multiple.real.2}
\end{table}

\section{DISCUSSION}

In this paper, for both global and multiple testing,  the precision matrix $\Ome=\Sig^{-1}$ of the covariates is assumed to be sparse and unknown. Node-wise regression among the covariates is used to learn the covariance structure in constructing  the debiased estimator. However, if the prior knowledge of $\Ome=\bold{I}$ is available, the algorithm can be simplified greatly. Specifically, instead of incorporating the Lasso estimators as in (\ref{lasso}), we let $v_j = \hat{W}^{-1}\bx_j$ and $\tau_j = \|v_j\|_n/\langle v_j,\bx_j\rangle$ for each $j=1,...,p$. The theoretical properties of the resulting global testing and multiple testing procedures still hold, while the computational efficiency  is improved dramatically. However, from  our theoretical analysis,  even  with the knowledge of  $\Ome=\bold{I}$, the theoretical requirement for the model sparsity ($k=o(\sqrt{n}/\log^3p)$ in the Gaussian case and $k=o(\sqrt{n}/\log^{5/2}p)$ in the bounded case)  cannot be relaxed due to the nonlinearity of the problem.

Sample splitting was used in this paper for theoretical purpose. This is different from  other works on inference in high-dimensional linear/logistic regression models, including  \cite{ingster2010detection}, \cite{van2014asymptotically}, \cite{mukherjee2015hypothesis} and \cite{javanmard2019false}, where sample splitting is not needed. However, as we discussed throughout the paper, the assumptions and the alternatives that we considered are different from those previous papers. In the case of high-dimensional logistic regression model,  a sample splitting procedure seems unavoidable under the current framework of our technical analysis without making  additional strong structural assumptions such as the sparse inverse Hessian matrices used in \cite{van2014asymptotically} or the weakly correlated design matrices used in \cite{mukherjee2015hypothesis}. Our simulations showed that the sample splitting is actually not needed in order for our proposed methods to perform well. It is of interest to develop technical tools that can eliminate  sample splitting in inference for high dimensional logistic regression models.

As mentioned in the introduction, the logistic regression model can be viewed as a special case of the single index model $y=f(\beta^\top x)+\epsilon$ where $f$ is a known transformation function \citep{yang2015sparse}. Based on our analysis, it is clear that the theoretical results are not limited to the sigmoid transfer function. In fact, the proposed methods can be applied to a wide range of transformation functions satisfying the following conditions: (C1) $f$ is continuous and for any $u\in \R$, $0<f(u)<1$; (C2) for any $u_1,u_2\in \R$, there exists a constant $L>0$ such that $|\dot{f}(u_1)-\dot{f}(u_2)|\le L|u_1-u_2|$; and (C3) for any constant $C>0$, there exists $\delta>0$ such that for any $|u| \le C$, $ \dot{f}(u) \ge \delta$.
Examples include but are not limited to the following function classes
\begin{itemize}
	\item \emph{Cumulative density functions:} $f(x) = P(X\le x)$ for some continuous random variable $X$ supported on $\R$. In particular, when $X\sim N(0,1)$, the resulting model becomes the probit regression.
	\item \emph{Affine hyperbolic tangent functions:} $f(x) = \frac{1}{2}{\tanh(ax+b)+\frac{1}{2}}$ for some parameter $a,b\in \R$. In particular, $(a,b)=(1,0)$ corresponds to $f(x) = {e^x}/({1+e^x})$.
	\item \emph{Generalized logistic functions:} $f(x) = (1+e^{-x})^{-\alpha}$ for some $\alpha>0$.
\end{itemize}

Besides the problems we considered  in this  paper, it is also of interest to construct confidence intervals for functionals of the regression coefficients, such as $\|\beta\|_1$, $\|\beta\|_2$, or $\theta^\top\beta$ for some given loading vector $\theta$. In modern statistical machine learning, logistic regression is considered as an efficient classification method \citep{abramovich2018high}. In practice, a predicted label with an uncertainty assessment is usually preferred. Therefore, another important problem is the construction of predictive intervals of the conditional probability $\pi^*$ associated with a given predictor $X^*$. These problems are related to the current work and are left for future investigations. 

\section{PROOFS OF THE MAIN THEOREMS}

In this section, we prove Theorems \ref{null.dis}, Theorem \ref{global.lower} and Theorem \ref{fdr} in the paper. The proofs of other results, including Theorems 3 and 5, Proposition 1 and the technical lemmas, are given in our Supplementary Materials. 

\paragraph{Proof of Theorem \ref{null.dis}} Define $F_{jj}=\E[\eta_{ij}^2/\dot{f}(u_i)]$. Under $H_0$, $F_{jj}=4\E[\eta_{ij}^2]=4/\omega_{jj}$, and by (A3), $c<F_{jj}<C$ for $j=1,...,p$ and some constant $C\ge c>0$. Define statistics
\[
\tilde{M}_j = \frac{\langle v_j,\epsilon\rangle}{\|v_j\|_n}, \quad\text{ and }\quad \check{M}_j = \frac{\sum_{i=1}^n\eta_{ij}\epsilon_i/\dot{f}(u_i)}{\sqrt{nF_{jj}}},\quad j=1,...,p.
\]
and $\tilde{M}_n = \max_{j}\tilde{M}_j^2, \check{M}_n = \max_{j}\check{M}_j^2$. The following lemma shows that $\tilde{M}_n$ and therefore $\check{M}_n$ are good approximations of $M_n$.
\bel \label{event2}
Under the condition of Theorem 1, the following events 
\begin{align*}
B_1 = \bigg\{ |\tilde{M}_n-\check{M}_n| =o(1) \bigg\},\quad \quad B_2= \bigg\{ |\tilde{M}_n-M_n| =o\bigg( \frac{1}{{\log p}} \bigg) \bigg\},
\end{align*}
hold with probability at least $1-O(p^{-c})$ for some constant $c>0$.
\eel 
It follows that under the event $B_1\cap B_2$, let $y_p = 2\log p-\log \log p+x$ and  $\epsilon_n=o(1)$, we have
\[
P_{\theta}\big( \check{M}_n \le y_p-\epsilon_n \big)\le P_{\theta}\big( {M}_n\le y_p\big)\le P_{\theta}\big( \check{M}_n\le y_p+\epsilon_n \big)
\]
Therefore it suffices to prove that for any $t\in\R$, as $(n,p)\to \infty$,
\beq
P_{\theta}\big( \check{M}_n\le y_p \big) \to  \exp\bigg( -\frac{1}{\sqrt{\pi}}\exp (-x/2)\bigg).
\eeq 
Now define $\hat{M}_j = \frac{\sum_{i=1}^n \hat{Z}_{ij}}{\sqrt{nF_{jj}}}, j=1,...,p.$
where $\hat{Z}_{ij}=v^0_{ij}\epsilon_i 1\{ |v^0_{ij}\epsilon_i| \le \tau_n\}- \E [v^0_{ij}\epsilon_i 1\{ |v^0_{ij}\epsilon_i| \le \tau_n\}]$ for $\tau_n = \log (p+n)$, $v^0_{ij}= \eta_{ij}/\dot{f}(u_i)$ and $\hat{M}_n = \max_{j}\hat{M}_j^2$. The following lemma states that $\hat{M}_n$ is close to $\check{M}_n$.
\bel \label{hatM.lem}
Under the condition of Theorem 1, $|\check{M}_n-\hat{M}_n| = o(1)$ with probability at least $1-O(p^{-c})$  for some constant $c>0$.
\eel

By Lemma \ref{hatM.lem}, it suffices to prove that for any $t\in\R$, as $(n,p)\to \infty$,
\beq \label{25}
P_{\theta}\big( \hat{M}_n\le y_p \big) \to  \exp\bigg( -\frac{1}{\sqrt{\pi}}\exp (-x/2)\bigg).
\eeq 
To prove this, we need the classical Bonferroni inequality.
\bel \label{bonferroni}
\textbf{\textup{(Bonferroni inequality)}} Let $B = \cup_{t=1}^p B_t$. For any integer $k<p/2$, we have
\beq
\sum_{t=1}^{2k}(-1)^{t-1}A_t \le P(B) \le \sum_{t=1}^{2k-1}(-1)^{t-1}A_t,
\eeq
where $A_t = \sum_{1\le i_1 <...<i_t \le p} P(B_{i_1}\cap ... \cap B_{i_t})$.
\eel
By Lemma \ref{bonferroni}, for any integer $0<q<p/2$,
\begin{align} \label{26}
\sum_{d=1}^{2q}(-1)^{d-1} \sum_{1\le j_1 <...<j_d \le p} P_\theta\bigg( \bigcap_{k=1}^d A_{j_k} \bigg) &\le P_\theta \bigg( \max_{1\le j\le p} \hat{M}_j^2 \ge y_p \bigg) \nonumber \\
& \le \sum_{d=1}^{2p-1}(-1)^{d-1}\sum_{1\le j_1 <...<j_d \le p} P_\theta\bigg( \bigcap_{k=1}^d A_{j_k} \bigg),
\end{align}
where $A_{j_k} = \{\hat{M}_{j_k}^2\ge y_p\}$. Now let $w_{i_j}=\hat{Z}_{ij}/\sqrt{F_{jj}}$ for $j=1,...,p$, and $\bold{W}_i=(w_{i,j_1},...,w_{i,j_d})^\top $ for $1\le i\le n$. Define $\| \bold{a}\|_{\min} = \min_{1\le i\le d}|a_i|$ for any vector $\bold{a}\in \R^d$. Then we have
\[
P_\theta\bigg( \bigcap_{k=1}^d A_{j_k} \bigg) = P_\theta \bigg(\bigg\| n^{-1/2}\sum_{i=1}^n \bold{W}_i \bigg\|_{\min} \ge y^{1/2}_p  \bigg).
\] 
Then it follows from Theorem 1.1 in \cite{zaitsev1987gaussian} that 
\begin{align}\label{27}
P_\theta \bigg(\bigg\| n^{-1/2}\sum_{i=1}^n \bold{W}_i \bigg\|_{\min} \ge y^{1/2}_p  \bigg) &\le P_\theta \bigg( \| \bold{N}_d\|_{\min} \ge y_p^{1/2}-\epsilon_n(\log p)^{-1/2} \bigg) \nonumber \\
&\quad+ c_1d^{5/2}\exp \bigg\{-\frac{n^{1/2}\epsilon_n}{c_2d^3\tau_n(\log p)^{1/2}}  \bigg\},
\end{align}
where $c_1>0$ and $c_2>0$ are constants, $\epsilon_n \to 0$ which will be specified later, and $\bold{N}_d = (N_{m_1},...,N_{m_d})$ is a normal random vector with $\E(\bold{N}_d) = 0$ and $\text{cov}(\bold{N}_d) = \text{cov}(\bold{W}_1)$. Here $d$ is a fixed integer that does not depend on $n,p$. Because $\log p = o(n^{1/5})$, we can let $\epsilon_n \to 0$ sufficiently slow, say, $\epsilon_n = \sqrt{{\log^5p}/{n}}$, so that for any large $c>0$,
\beq\label{28}
c_1 d^{5/2} \exp \bigg\{ -\frac{n^{1/2}\epsilon_n}{c_2 d^3 \tau_n (\log p)^{1/2}}\bigg\} = O(p^{-c}).
\eeq 
Combining (\ref{26}), (\ref{27}) and (\ref{28}), we have
\begin{align} \label{29}
P_\theta \bigg( \max_{1\le j\le p} \hat{M}_j^2 \ge y_p \bigg) \le  \sum_{d=1}^{2p-1}(-1)^{d-1}\sum_{1\le j_1 <...<j_d \le p} P_\theta \bigg( \| \bold{N}_d\|_{\min} \ge y_p^{1/2}-\epsilon_n(\log p)^{-1/2} \bigg) +o(1).
\end{align}
Similarly, one can derive
\beq \label{30}
P_\theta \bigg( \max_{1\le j\le p} \hat{M}_j^2 \ge y_p \bigg) \ge  \sum_{d=1}^{2p}(-1)^{d-1}\sum_{1\le j_1 <...<j_d \le p} P_\theta \bigg( \| \bold{N}_d\|_{\min} \ge y_p^{1/2}+\epsilon_n(\log p)^{-1/2} \bigg) +o(1).
\eeq
Now we use the following lemma from \cite{xia2018two}.
\bel \label{tony2013}
For any fixed integer $d\ge 1$ and real number $t\in \R$,
\[
\sum_{1\le j_1 <...<j_d \le p} P_\theta \bigg( \| \bold{N}_d\|_{\min} \ge y_p^{1/2}\pm\epsilon_n(\log p)^{-1/2} \bigg) =\frac{1}{d!}\bigg( \frac{1}{\sqrt{\pi}}\exp (-t/2) \bigg)^d(1+o(1)).
\]
\eel
It then follows from the above lemma, (\ref{29}) and (\ref{30}) that
\begin{align*}
\limsup_{n,p\to \infty} P_\theta \bigg( \max_{1\le j\le p} \hat{M}_j^2 \ge y_p \bigg) &\le  \sum_{d=1}^{2p}(-1)^{d-1}\frac{1}{d!}\bigg( \frac{1}{\sqrt{\pi}}\exp (-t/2) \bigg)^d, \\
\liminf_{n,p\to \infty} P_\theta \bigg( \max_{1\le j\le p} \hat{M}_j^2 \ge y_p \bigg) &\ge  \sum_{d=1}^{2p-1}(-1)^{d-1}\frac{1}{d!}\bigg( \frac{1}{\sqrt{\pi}}\exp (-t/2) \bigg)^d, 
\end{align*}
for any positive integer $p$. By letting $p\to \infty$, we obtain (\ref{25}) and the proof is complete.
\qed

\paragraph{Proof of Theorem \ref{global.lower}.} The proof essentially follows from the general Le Cam's method described in Section 7.1 of \cite{baraud2002non}. The key elements can be summarized as the following lemma that reduces the lower bound problem to calculation of the total variation distance between two posterior distributions.

\bel \label{baraud}
Let $\HH_1$ be some subset in an $\ell_2$ bounded Hilbert space and $\rho$ some positive number. Let $\mu_\rho$ be some probability measure on $\HH_1 = \{ \theta\in \Theta,\|\theta\|=\rho\}.$ Set $P_{\mu_\rho}= \int P_\theta d\mu_\rho(\theta)$, $P_0$ as the (posterior) distribution at the null, and denote by $\Phi_\alpha$ the level-$\alpha$ tests, we have
\[
\inf_{\Phi_\alpha}\sup_{\theta\in\HH_1}P_\theta(\Phi_\alpha=0)\ge \inf_{\Phi_\alpha} P_{\mu_\rho}(\Phi_\alpha=0) \ge 1-\alpha-TV(P_{\mu_\rho},P_0),
\]
where $TV(P_{\mu_\rho},P_0)$ denotes the total variation distance between $P_{\mu_\rho}$ and $P_0$.
\eel
Now since by definition $\rho^*(\Phi_\alpha,\delta,\Theta(k))\ge \rho^*(\Phi_\alpha,\delta,\Sig)$ for any $\Sig\in \Theta_{2}(k)$, by Lemma \ref{baraud}, it suffices to construct the corresponding $\HH_1$ for $\beta\in \Theta_\beta(k)$ and find a lower bound $\rho_1=\rho(\eta)$ such that 
\beq \label{tv.dis}
\forall \rho\le \rho_1 \quad \quad \inf_{\Phi_\alpha} P_{\mu_\rho}(\Phi_\alpha=0)\ge 1-\alpha-\eta = \delta.
\eeq
for fixed covariance $\Sig=\bold{I}$.
In this case, an upper bound for the $\chi^2$-divergence between $P_{\mu_\rho}$ and $P_0$, defined as $\chi^2(P_{\mu_\rho},P_0) = \int \frac{(dP_{\mu_\rho})^2}{dP_0} -1$, can be obtained by carefully constructing the alternative space $\HH_1$.
Since $TV(f,g) \le \sqrt{\chi^2(f,g)}$ (see p.90 of \cite{tsybakov2009introduction}),
it follows that $\inf_{\Phi_\alpha} P_{\mu_\rho}(\Phi_\alpha=0)\ge 1-\alpha-\sqrt{\chi^2(P_{\mu_\rho},P_0) }.$
By choosing $\rho_1=\rho(\eta)$ such that for any $\rho\le \rho_1$, $\chi^2(P_{\mu_\rho},P_0)\le\eta^2=(1-\alpha-\delta)^2$, we have (\ref{tv.dis}) holds. In the following, we will construct the alternative space $\HH_1$ and derive an upper bound of $\chi^2(P_{\mu_\rho},P_0)$ where $P_0$ corresponds to the null space $\HH_0$ defined at a single point $\beta=0$. We divide the proofs into two parts. Throughout, the design covariance matrix is chosen as $\Sig=\bold{I}$.

\paragraph{Step 1: Construction of $\HH_1$.} 
Firstly, for a set $M$, we define $\ell(M,n)$ as the set of all the $n$-element subsets of $M$. Let $[1:p] \equiv \{1,...,p\}$, so $\ell([1:p],k)$ contains all the $k$-element subsets of $[1:p]$. We define the alternative parameter space $\mathcal{H}_1= \big\{ \beta\in \R^p:  \beta_{j} = \rho1\{j\in I\} \text{ for } I \in \ell([1:p],k) \big\}.$
In other words, $\mathcal{H}_1$ contains all the $k$-sparse vectors $\beta(I)$ whose nonzero components $\rho$ are indexed by $I$. Apparently, for any $\beta \in \HH_1$, it follows $\|\beta\|_\infty=\rho$ and $\HH_1 \subseteq \Theta_1(k)$. 

\paragraph{Step 2: Control of $\chi^2(P_{\pi_{\HH_1}},P_0)$.} 
Let $\pi$ denote the uniform prior of the random index set $I$ over $\ell([1:p],k)$. This prior induces a prior distribution $\pi_{\HH_1}$ over the parameter space $\HH_1$. For $\{ {\bf 0}_p\}=\mathcal{H}_0$, the corresponding joint distribution of the data $\{(X_i,y_i)\}_{i=1}^n$ is
\[
f = \prod_{i=1}^n p(X_i,y_i) =\frac{1}{(2\pi)^{np/2}} \prod_{i=1}^n  \frac{1}{2}e^{- \|X_i\|_2^2/2}.
\]
Similarly, the posterior distribution of the samples over the prior $\pi_{\HH_1}$ is denoted as
\[
g = \prod_{i=1}^n \int_{\HH_1} p(X_i,y_i;\beta) \pi_{\HH_1} =\frac{1}{{p \choose k}}   \sum_{\beta\in \mathcal{H}_1} \prod_{i=1}^n p(X_i,y_i;\beta).
\]
As a result, we have the following lemma controlling $\chi^2(P_{\pi_{\HH_1}},P_0)=\chi^2(g,f)$.
\bel \label{chisq.lem}
Let $\rho^2 = \frac{1}{n}\log\big(1+\frac{p}{h(\eta)k^2}\big)$ where $h(\eta)=[\log(\eta^2+1)]^{-1}$ and $\eta = 1-\alpha-\delta$, then we have $\chi^2(g,f) \le (1-\alpha-\delta)^2.$
\eel
Combining Lemma \ref{baraud} and Lemma \ref{chisq.lem}, we know that for $\alpha,\delta>0$ and $\alpha+\delta<1$, if $\rho =  \sqrt{\frac{1}{n}\log\big(1+\frac{p}{h(\eta)k^2}\big)}$, then $\forall \rho'\le \rho, \inf_{\Phi_\alpha}\sup_{\beta\in\Theta(k):\|\beta\|_\infty \ge \rho'}P_\theta(\Phi_\alpha=0)\ge \delta.$
Therefore, it follows that
\beq \label{lower1}
\rho^*(\alpha,\delta,\Theta(k)) \ge \rho^*(\alpha,\delta,\bold{I}) \gtrsim   \sqrt{\frac{1}{n}\log \bigg( 1+\frac{p}{k^2}\bigg)}.
\eeq
Lastly, note that for the above chosen $\rho$, $\mathcal{H}_1\subset \Theta_1(k)\cap  \{ \beta\in \R^p: \|\beta\|_2\lesssim (n^{1/4}\log p)^{-1} \}$ when $k \lesssim \min\{p^\gamma, \sqrt{n}/\log^3 p\}$ for some $0< \gamma <1/2$. This completes the proof.
\qed

\paragraph{Proof of Theorem \ref{fdr}.} The proof follows similar arguments of the proof of Theorem 3.1 in \cite{javanmard2019false}. We first consider the case when $\hat{t}$, given by (\ref{t.hat.fdr}), does not exist.
In this case, $\hat{t}=\sqrt{2\log p}$ and we consider the event $\Omega_0=\{ \sum_{j\in\HH_0} I\big(|M_j|\ge \sqrt{2\log p} \big)\ge 1 \}$ that there are at least one false positive. In order to show the FDR/FDP can be controlled in this case, we show that
\beq \label{prob.conv}
P_\theta( \Omega_0) \to 0,\quad \text{as $(n,p)\to\infty$.}
\eeq
Note that for $j\in \HH_0$, we have $M_j=\frac{\check{\beta}_j}{\tau_j}=\frac{\langle v_j,\epsilon\rangle}{\|v_j\|_n}+\frac{\langle v_j,Re \rangle}{\|v_j\|_n}-\frac{\langle v_j,\bf{h}_{-j} \rangle_n}{\|v_j\|_n}.$
Then 
\begin{align} \label{fdr.out}
P_\theta(\Omega_0 )&\le P_\theta\bigg( \sum_{j\in\HH_0} I\bigg(\frac{\langle v_j,\epsilon\rangle}{\|v_j\|_n}+\frac{\langle v_j,Re \rangle}{\|v_j\|_n}-\frac{\langle v_j,\bf{h}_{-j} \rangle_n}{\|v_j\|_n}\ge \sqrt{2\log p} \bigg)\ge 1 \bigg)\nonumber\\
&\quad+P_\theta\bigg( \sum_{j\in\HH_0} I\bigg(\frac{\langle v_j,\epsilon\rangle}{\|v_j\|_n}+\frac{\langle v_j,Re \rangle}{\|v_j\|_n}-\frac{\langle v_j,\bf{h}_{-j} \rangle_n}{\|v_j\|_n}\le -\sqrt{2\log p} \bigg)\ge 1 \bigg).
\end{align}
For any $\epsilon>0$, we can bound the first term by
\begin{align*}
&P_\theta\bigg( \sum_{j\in\HH_0} I\bigg(\frac{\langle v_j,\epsilon\rangle}{\|v_j\|_n}+\frac{\langle v_j,Re \rangle}{\|v_j\|_n}-\frac{\langle v_j,\bf{h}_{-j} \rangle_n}{\|v_j\|_n}\ge \sqrt{2\log p} \bigg)\ge 1 \bigg)\\
&=P_\theta\bigg( \sum_{j\in\HH_0} I\bigg(\tilde{M}_j\ge \sqrt{2\log p}+\frac{\langle v_j,\bf{h}_{-j} \rangle_n}{\|v_j\|_n}-\frac{\langle v_j,Re \rangle}{\|v_j\|_n} \bigg)\ge 1 \bigg)\\
&\le P_\theta\bigg( \sum_{j\in\HH_0} I\bigg(\tilde{M}_j\ge \sqrt{2\log p}-\epsilon\bigg)\ge 1\bigg)+P_\theta\bigg(\max_{j\in \HH_0} \bigg|\frac{\langle v_j,\bf{h}_{-j} \rangle_n}{\|v_j\|_n}-\frac{\langle v_j,Re \rangle}{\|v_j\|_n}\bigg|\ge\epsilon \bigg)\\
&\le p\max_{j\in\HH_0} P_\theta\bigg( \tilde{M}_j\ge \sqrt{2\log p}-\epsilon\bigg)+P_\theta\bigg(\max_{j\in \HH_0} \bigg|\frac{\langle v_j,\bf{h}_{-j} \rangle_n}{\|v_j\|_n}-\frac{\langle v_j,Re \rangle}{\|v_j\|_n}\bigg|\ge\epsilon \bigg)
\end{align*}
By the proof of Lemma 1, we know that $P_\theta\big( \max_{j\in \HH_0}\big|\frac{\langle v_j,\bf{h}_{-j} \rangle_n}{\|v_j\|_n}-\frac{\langle v_j,Re \rangle}{\|v_j\|_n}\big|\ge\epsilon \big)\to 0.$
In addition, for $j\in \HH_0$, $P_\theta\big( \tilde{M}_j\ge \sqrt{2\log p}-\epsilon\big)\le P_\theta\big( \check{M}_j\ge \sqrt{2\log p}-2\epsilon\big)+P_\theta(|\tilde{M}_j-\check{M}_j|\ge \epsilon),$ where $\max_{j\in \HH_0}P_\theta(|\tilde{M}_j-\check{M}_j|\ge \epsilon)= O(p^{-c})$ for some sufficiently large $c>0$. Now since $\check{M}_j=\frac{\sum_{i=1}^n\eta_{ij}\epsilon_i/\dot{f}(u_i)}{\sqrt{nF_{jj}}}$ where $\E \frac{\eta_{ij}\epsilon_i/\dot{f}(u_i)}{\sqrt{F_{jj}}}=0$ and $\text{Var}(\frac{\eta_{ij}\epsilon_i/\dot{f}(u_i)}{\sqrt{F_{jj}}})=1$, by Lemma 6.1 of \cite{liu2013gaussian}, we have $\sup_{0\le t\le 4\sqrt{\log p}}\big| \frac{P_\theta(|\check{M}_j|\ge t)}{G(t)}-1\big|\le C(\log p)^{-1}.$
Now let $t=\sqrt{2\log p}-2\epsilon$, we have
\[
P_\theta\bigg( \check{M}_j\ge \sqrt{2\log p}-2\epsilon\bigg)\le G(\sqrt{2\log p}-2\epsilon)+C\frac{G(\sqrt{2\log p}-2\epsilon)}{\log p}.
\]
Hence $p\max_{j\in\HH_0} P_\theta\bigg( \tilde{M}_j\ge \sqrt{2\log p}-\epsilon\bigg)\le CpG(\sqrt{2\log p}-2\epsilon)+O(p^{-c}),$ which goes to zero as $(n,p)\to\infty$. By symmetry, we know that the second term in (\ref{fdr.out}) also goes to 0. Therefore we have proved (\ref{prob.conv}).

Now consider the case when $0\le \hat{t}\le b_p$ holds. We have
\[
\text{FDP}_\theta(\hat{t})= \frac{\sum_{j\in \mathcal{H}_0} I\{ |M_{j}| \ge \hat{t}\}}{\max \big\{\sum_{j=1}^p I\{ |M_{j}| \ge \hat{t}\},1   \big\}}\le \frac{p_0G(\hat{t}) }{\max \big\{\sum_{j=1}^p I\{ |M_{j}| \ge \hat{t}\},1   \big\}}(1+A_p),
\]
where $A_p=\sup_{0\le t\le b_p}\big|\frac{\sum_{j\in \mathcal{H}_0} I\{ |M_{j}| \ge t\}}{p_0G(t)} -1\big|.$
Note that by definition $\frac{p_0G(\hat{t}) }{\max \big\{\sum_{j=1}^p I\{ |M_{j}| \ge \hat{t}\},1   \big\}}\le \frac{p_0\alpha}{p}.$
The proof is complete if $A_p\to 0$ in probability, which has been shown by Proposition 1.
\qed

\section*{FUNDING}
This research was supported by NIH grants R01CA127334 and R01GM123056. 

\section*{SUPPLEMENTARY MATERIALS}

In the online Supplemental Materials, we prove Theorem 3, 5, Proposition 1, and the technical lemmas. The technical results and simulations concerning the two-sample tests discussed in Section 4 are also included.

\bibliographystyle{chicago}
\bibliography{reference}

\newpage

\title{Supplement to ``Global and Simultaneous Hypothesis Testing for High-Dimensional Logistic Regression Models"}
\author{Rong Ma$^1$, T. Tony Cai$^2$ and Hongzhe Li$^1$ \\
	Department of Biostatistics, Epidemiology and Informatics$^1$\\
	Department of Statistics$^2$\\
	University of Pennsylvania\\
	Philadelphia, PA 19104}
\date{}
\maketitle
\thispagestyle{empty}

\begin{abstract}
	In this Supplementary Material we prove Theorem 3, 5 and Proposition 1 in the main paper and the technical lemmas. The technical and simulation  results of the two-sample tests discussed in Section 4 of the main paper are included in the appendix.
\end{abstract}

\setcounter{section}{0}

\section{Proofs of Main Results}

\setcounter{lemma}{6}

\subsection{Proof of Proposition 1}

By similar argument as in Lemma 1, we can prove the following lemma.
\bel \label{event4}
Assume (A2) (A3) and (A4), $k=o(\sqrt{n}/\log^{5/2} p)$, then 
\begin{align*}
\max_{j\in \HH_0}|\tilde{M}_j-\check{M}_j| =o\bigg( \frac{1}{\sqrt{\log p}} \bigg), \quad  \max_{j \in \HH_0}|\tilde{M}_j-M_j| =o\bigg( \frac{1}{\sqrt{\log p}} \bigg),
\end{align*}
hold with probability at least $1-O(p^{-c})$ for some constant $c>0$.
\eel
For (20), by Lemma 6.1 in \cite{liu2013gaussian}, we have
\beq \label{5.13}
\max_{1\le j\le p}\sup_{0\le t\le 4\sqrt{\log p}}\bigg| \frac{P_\theta(|\check{M}_j| \ge t)}{G(t)}-1 \bigg|\le C(\log p)^{-2-\gamma_1}
\eeq
for some constant $0<\gamma_1<1/2$. So (20) follows from Lemma \ref{event4}, and the fact that $G(t+o(1/\sqrt{\log p}))/G(t) = 1+o(1)$ uniformly in $0\le t\le \sqrt{2\log p}$.

For (21), 
it suffices to show that
\beq \label{5.14}
\sup_{0\le t\le b_{p}} \bigg|  \frac{\sum_{j\in \HH_0} I\{|\check{M}_j| \ge t \}}{p_0G(t)}-1 \bigg|\rightarrow 0\quad \text{ in probability.}
\eeq
Let $z_0<z_1<...<z_{d_p}\le 1$ and $t_i = G^{-1}(z_i)$, where $z_0 = G(b_p)$, $z_i = c_p/p+c_p^{2/3}e^{i^{\delta}}/p$ with $c_p=pG(b_p)$, and $d_p = [\log ((p-c_p)/c_p^{2/3})]^{1/\delta}$ and $0<\delta<1$, which will be specified later. We have $G(t_i)/G(t_{i+1}) = 1+o(1)$ uniformly in $i$, and $t_0/\sqrt{2\log(p/c_p)}=1+o(1)$. Note that uniformly for $1\le j\le m$, $G(t_i)/G(t_{i-1})\to 1$ as $p\to \infty$. The proof of (\ref{5.14}) reduces to show that
\beq \label{5.15}
\max_{0\le i\le d_{p}} \bigg|  \frac{\sum_{j\in \HH_0} I\{|\check{M}_j| \ge t_i \}}{p_0G(t_i)}-1 \bigg|\rightarrow 0
\eeq
in probability. In fact, for each $\epsilon >0$, we have
\begin{align*}
&P_\theta\bigg(   \max_{0\le i\le d_{p}} \bigg| \frac{\sum_{j\in \HH_0} [I\{|\check{M}_j| \ge t_i \}-G(t_i) ]}{p_0 G(t_i)} \bigg|\ge \epsilon \bigg)  \le \sum_{j=0}^{d_p} P_\theta\bigg(   \bigg| \frac{\sum_{j\in \HH_0} [I\{|\check{M}_j| \ge t_i \}-G(t_i) ]}{p_0 G(t_i)} \bigg|\ge \epsilon/2 \bigg).
\end{align*}
Set $I(t) = \frac{\sum_{j\in \HH_0} [I\{|\check{M}_j| \ge t \}-P_\theta(|\check{M}_j| \ge t)]}{p_0 G(t)}.$
By Markov's inequality $P_\theta(|I(t_i) |\ge \epsilon /2) \le \frac{\E [I(t_i)]^2 }{\epsilon^2/4},$
and it suffices to show $\sum_{j=0}^{d_p} \E [I(t_i)]^2=o(1)$.
To see this, by (\ref{5.13}),
\begin{align*}
\E I^2(t) &= \frac{\sum_{j\in \HH_0} [P_\theta(|\check{M}_j| \ge t)-P_\theta^2(|\check{M}_j| \ge t)]}{p_0^2G^2(t)}\\
&\quad + \frac{\sum_{j,k\in \HH_0, k\ne j} [P_\theta(|\check{M}_j| \ge t,|\check{M}_k| \ge t)-P_\theta(|\check{M}_j| \ge t)P_\theta(|\check{M}_k| \ge t)]}{p_0^2G^2(t)}\\
& \le \frac{C}{p_0G(t)} + \frac{1}{p_0^2}\sum_{(j,k)\in \mathcal{A}(\epsilon): j,k\in \HH_0} \frac{P_\theta(|\check{M}_j| \ge t,|\check{M}_k| \ge t)}{G^2(t)}\\
&\quad+  \frac{1}{p_0^2}\sum_{(j,k)\in \mathcal{A}(\epsilon)^c: j,k\in \HH_0} \bigg[\frac{P_\theta(|\check{M}_j| \ge t,|\check{M}_k| \ge t)}{G^2(t)} -1\bigg]\\
&= \frac{C}{p_0G(t)} + I_{11}(t) + I_{12}(t).
\end{align*}
For $(j,k)\in \mathcal{A}(\epsilon)^c$ with $j,k\in \HH_0$, applying Lemma 6.1 in \cite{liu2013gaussian}, we have $I_{12}(t) \le C(\log p)^{-1-\xi}$
for some $\xi>0$ uniformly in $0<t<\sqrt{2\log p}$. By Lemma 6.2 in \cite{liu2013gaussian}, for $(j,k)\in \mathcal{A}(\epsilon)$ with $j,k\in \HH_0$, we have
\[
P_\theta(|\check{M}_j| \ge t,|\check{M}_k| \ge t) \le C(t+1)^{-2}\exp \bigg(-\frac{t^2}{1+|\rho_{jk}|} \bigg).
\]
So that
\[
I_{11}(t) \le C \frac{1}{p_0^2}\sum_{(j,k)\in \mathcal{A}(\epsilon): j,k\in \HH_0}(t+1)^{-2}\exp \bigg(-\frac{t^2}{1+|\rho_{jk}|} \bigg)G^{-2}(t) \le C \frac{1}{p_0^2}\sum_{(j,k)\in \mathcal{A}(\epsilon): j,k\in \HH_0} [G(t)]^{-\frac{2|\rho_{jk}|}{1+|\rho_{jk}|}}.
\]
Note that for $0\le t\le b_p$, we have $G(t) \ge G(b_p) = c_p/p$, so that by assumption (A5) it follows that for some $\epsilon,q>0$,
\[
I_{11}(t) \le C\sum_{(j,k)\in \mathcal{A}(\epsilon): j,k\in \HH_0} p^{\frac{2|\rho_{jk}|}{1+|\rho_{jk}|}+q-2} = O(1/(\log p)^2).
\]
By the above inequalities, we can prove (\ref{5.15}) by choosing $0<\delta<1$ so that
\begin{align*}
\sum_{i=0}^{d_p}\E[I(t_i)]^2 &\le C\sum_{i=0}^{d_p}(pG(t_i))^{-1}+Cd_p[(\log p)^{-1-\delta}+(\log p)^{-2}]\\
&\le C\sum_{i=0}^{d_p} \frac{1}{c_p+c_p^{2/3}e^{i^{\delta}}}+o(1)\\
&=o(1).
\end{align*}
\qed

\subsection{Proof of Theorem 3}

Define $M'_j = \tau_j^{-1}( \check{\beta}_j-\beta_j)$,
and $M'_n = \max_j (M'_j)^2$, we have $-\beta_j/\tau_j = M'_j-M_j$.
Thus
\beq \label{dec.bnd2}
{\beta^2_j}/{\tau^2_j} \le 2(M'_j)^2+2M_j^2, \quad \text{for all $j$,}
\eeq
and
\beq \label{dec.bound}
\max_j {\beta^2_j}/{\tau^2_j} \le 2M'_n+2M_n.
\eeq
The main idea for proving Theorem 3 is that, in order to show that $M_n$ is ``large", we show that $M'_n$ is ``small" while $\max_j \beta_j^2/\tau_j^2$ is ``large" under the condition of Theorem 3. In the following, we consider the Gaussian design and the bounded design separately.
For the Gaussian design, we divide the proof into two parts.
\paragraph{Gaussian Design, Case 1. $\|\beta\|_2\lesssim (\log p)^{-1/2}$.} In this case, $\beta^\top X_i$ are i.i.d. $N(0,\beta^\top\Sig\beta)$. By Lemma 6 in \cite{tony2014two}, we have
\beq \label{betax.bnd}
P_\theta\bigg(\max_{1\le i\le n}|\beta^\top X_i| \ge \|\beta\|_2\sqrt{2\lambda_{\max}(\Sig)\log p}\bigg)=O(p^{-c}),
\eeq  
then ({A4}), or $P_\theta(\max_{1\le i\le n}|\beta^\top X_i| \le c) \to 1$ for some constant $c>0$, holds. Consequently, the following lemma can be established by similar arguments as the proof of Lemma 1.

\bel \label{events3}
Under the condition of Theorem 3, suppose (A4) hold, then
\beq \label{m'jbound}
P_\theta\big( |M'_j|\ge \sqrt{C_0\log p} \big) = O(p^{-c})
\eeq
for some constants $C_0, c>0$. 
\eel

By Lemma \ref{events3}, we have
\beq \label{mnbound1}
P_\theta\big( M'_n\ge C_0{\log p} \big) = O(p^{-c})
\eeq
for some $C_0,c>0$. 
On the other hand, to bound $\tau_j$, we start with the inequality
\[
\frac{\|\hat{\eta}_j\|_2}{\langle \hat{\eta}_j, \bx_j \rangle} \le \frac{C_2}{\sqrt{n}}
\]
obtained as (\ref{zhangzhang}) in the proof of Lemma 1.
By (A4), there exists some constant $0<\kappa<1$ such that $\kappa<|f(u_i)|<1-\kappa$ with high probability. Then it follows that 
\[
1-\dot{f}(\hat{u}_i) \le \xi \dot{f}(\hat{u}_i), \quad\quad\text{where $\xi_1 = \frac{1-\kappa+\kappa^2}{\kappa(1-\kappa)}$}.
\]
Thus, since
\[
\|v_j\|_n-\|\hat{\eta}_j\|_2 \le \sqrt{\sum_{i=1}^n(\dot{f}(\hat{u}_i) - \dot{f}^2(\hat{u}_i))v^2_{ij}} \le \sqrt{\xi_1\sum_{i=1}^n \dot{f}^2(\hat{u}_i)v^2_{ij}} = \xi_1^{1/2}\|\hat{\eta}_j\|_2,
\]
we have, with probability at least $1-O(p^{-c})$,
\beq \label{tau.bnd}
\tau_j = \frac{\|v_j\|_{n}}{|\langle v_j,\bold{x}_j \rangle_n|} \le (1+\xi_1^{1/2}) \frac{\|\hat{\eta}_j\|_2}{|\langle \hat{\eta}_j,\bold{x}_j \rangle|}  \le C_2\frac{1+\xi_1^{1/2}}{\sqrt{n}} = \frac{C_3}{\sqrt{n}},
\eeq
for some constant $C_3>0$.
Therefore, since $\|\beta\|_\infty\ge c_2\sqrt{\log p/n}$,
\beq \label{beta.tau.bd}
\max_j \beta^2_j/\tau^2_j \ge c_2^2 \frac{\log p}{n} \cdot C_3^{-2}n = C_4\log p
\eeq
with probability converging to 1. In particular, when $c_2$ is chosen such that the constant $C_4-2C_0\ge4$, then under $H_1$, combining (\ref{dec.bound}) (\ref{mnbound1}) and (\ref{beta.tau.bd}), we have $P_{\theta}\big( \Phi_\alpha(M_n)=1\big) \to 1$ as  $(n,p)\to \infty$. 

\paragraph{Gaussian Design, Case 2. $\|\beta\|_2\gtrsim (\log p)^{-1/2}$.} In this case, we have
\beq \label{beta.bnd}
\|\beta\|_\infty \ge \sqrt{\|\beta\|_2^2/k} \gtrsim (k\log p)^{-1/2}.
\eeq
By (\ref{betax.bnd}), with probability at least $1-O(n^{-c})$,
\beq \label{fdot.bnd}
\min_{1\le i\le n} \dot{f}(u_i) \ge \frac{\exp(\|\beta\|_2\sqrt{2\lambda_{\max}(\Sig)\log n})}{(1+\exp(\|\beta\|_2\sqrt{2\lambda_{\max}(\Sig)\log n}))^2} \ge \frac{1}{4e^{\|\beta\|_2\sqrt{2\lambda_{\max}(\Sig)\log n}}}.
\eeq
Let
\[
L(n) = {e^{-\|\beta\|_2\sqrt{2\lambda_{\max}(\Sig)\log n}}}/4,
\]
it follows that with probability at least $1-O(n^{-c})$,
\[
1-\dot{f}(\hat{u}_i) \le \xi_2 \dot{f}(\hat{u}_i), \quad\quad\text{where $\xi_2 = \frac{1-L(n)}{L(n)}$}.
\]
Thus, with probability at least $1-O(n^{-c})$
\beq \label{tau.bnd}
\tau_j = \frac{\|v_j\|_{n}}{|\langle v_j,\bold{x}_j \rangle_n|} \le (1+\xi_2^{1/2}) \frac{\|\hat{\eta}_j\|_2}{|\langle \hat{\eta}_j,\bold{x}_j \rangle|}  \le C_2\frac{1+\xi_2^{1/2}}{\sqrt{n}} \le \frac{C_3e^{\|\beta\|_2\sqrt{0.5\lambda_{\max}(\Sig)\log n}}}{\sqrt{n}},
\eeq
for some constant $C_3>0$. Therefore, for $j=\arg\max |\beta_j|$, plug in (\ref{beta.bnd}) and $k=o(\sqrt{n}/\log^3p)$, we have
\beq
\beta^2_j/\tau^2_j \gtrsim \frac{n}{k\log p}e^{-\|\beta\|_2\sqrt{2\lambda_{\max}(\Sig)\log n}} \ge C_4 {\sqrt{n}\log^2p}e^{-\|\beta\|_2\sqrt{2\lambda_{\max}(\Sig)\log n}}
\eeq
with probability at least $1-O(n^{-c})$. Observe that as long as $\|\beta\|_2 \le C'\sqrt{\log n}$ for $C'=(2\sqrt{2\lambda_{\max}(\Sig)})^{-1}$ (which is true since by assumption $\log \log p\lesssim r\log n$ and $\|\beta\|_2\le C\log\log p/\sqrt{\log n}$ for some $C\le  (2r\sqrt{2\lambda_{\max}(\Sig)})^{-1}$), we have
\beq \label{1}
\beta^2_j/\tau^2_j \ge C_4 \log^2 p
\eeq
with probability at least $1-O(n^{-c})$.

Now we show that for the same $j=\arg\max |\beta_j|$,
\beq \label{mnbound}
P_\theta\big( (M'_j)^2\ge C_0{\log p} \big) = O(n^{-c})
\eeq
for some $C_0,c>0$. This can be established by the following lemma.

\bel \label{thm3.lem2}
Under the condition of Theorem 3, if $\|\beta\|_2\gtrsim (\log p)^{-1/2}$, then for any $j=1,...,p$,
\beq 
P_\theta\big( M'_j\ge C_1\sqrt{\log p} \big) = O(n^{-c})
\eeq
for some constants $C_1,c>0$.
\eel

Therefore, by (\ref{dec.bnd2}) (\ref{1}) and (\ref{mnbound}), we have
\[
M_n\ge M_j^2\ge \frac{1}{2}C_4 \log^2 p-C_0{\log p}
\]
with probability at least $1-O(n^{-c})$.
Thus $P_{\theta}\big( \Phi_\alpha(M_n)=1\big) \to 1$ as  $n\to \infty$.

\paragraph{Bounded Design.} The proof under the bounded design follows the same argument as the Case 1 of the Gaussian design, thus is omitted.
\qed

\subsection{Proof of Theorem 5}

By (20) in Proposition 1, let $t=\hat{t}_{FDV}$, it follows that as $(n,p)\to\infty$,
\beq
\sup_{j\in \mathcal{H}_0} \bigg| \frac{P_\theta(|M_{j}|\ge \hat{t}_{FDV})}{G(\hat{t}_{FDV})}-1  \bigg| \rightarrow 0,
\eeq
So that by noting that $G(\hat{t}_{FDV}) = r/p$, we have as $(n,p)\to\infty$,
\beq \label{fdv.bnd}
\bigg|\frac{\sum_{j\in \HH_0}P_\theta(|M_{j}|\ge \hat{t}_{FDV})}{r/p}-p_0\bigg| \to 0,
\eeq
which completes the proof of (23). To prove (24), it suffices to note that
\begin{align*}
\text{FWER}_\theta(t)=P_\theta\bigg( \sum_{j\in \HH_0}I(|M_j|\ge t)\ge 1 \bigg)=P_\theta\bigg(\bigcup_{j\in \HH_0}\{|M_j|\ge t  \}  \bigg)\le \sum_{j\in \HH_0}P_\theta(|M_j|\ge t),
\end{align*}
and the final result follows from (\ref{fdv.bnd}).
\qed

\section{Proofs of Technical Lemmas}

\paragraph{Proof of Lemma 1.} We start with the following lemma. In general, we will prove Lemma 1 under more general conditions posed in this lemma.
\bel \label{lem.var}
If one of the following two conditions holds,
\begin{enumerate}
	\item[(C1)] under Gaussian design, assume (A1) (A3) hold, $k = o(\sqrt{n}/\log^{3} p)$, and $\|X\beta\|_\infty \le c_2$ for some constant $c_2>0$;
	\item[(C2)] under the bounded design, assume (A2) (A3) (A4) hold, and $k = o(\sqrt{n}/\log^{5/2} p)$,
\end{enumerate}
then
\beq \label{variance.error}
\max_{1\le j\le p} \bigg|\frac{\|v_j\|_n}{\sqrt{n}}-F_{jj}^{1/2}   \bigg|=  o\bigg(  \frac{1}{\log p}\bigg)  
\eeq
in probability.
\eel
Lemma \ref{lem.var} can be established by combining results from Lemma \ref{events} and Lemma \ref{events.bounded} below, which provide some high probability bounds under the Gaussian and the bounded design, respectively.
\bel \label{events}
Under the Gaussian design, assume (A1) and (A3) hold, the following events
\begin{align*}
A_0 & =  \bigg\{ \|\hat{\beta}-\beta\|_1=O\bigg(k\sqrt{\frac{\log p}{{n}}}\bigg)\bigg\}, \\
A_1 & = \bigg\{ \max_{1\le j\le p}\frac{1}{n}\|X_{-j}(\hat{\gamma_j}-\gamma_j)\|_2^2 =O\bigg( k\frac{\log p}{n}\bigg)\bigg\},\\
A_2 & = \bigg\{ \max_{1\le j\le p} \|\hat{\gamma}_j-\gamma_j\|_1 =O\bigg(k\sqrt{\frac{\log p}{{n}}}\bigg)\bigg\},\\
A_3 & = \bigg\{ \max_{i,j}\big| {\hat{\eta}_{ij}}-{\eta_{ij}} \big| = O\bigg({\frac{k\log p}{\sqrt{n}}}\bigg)\bigg\} ,
\end{align*}
hold with probability at least $1-O(p^{-c})$ for some constant $c>0$. In addition, if $\|X\beta\|_\infty \le c_1$ for some constant $c_1>0$ and $k=o(n)$, the following events
\begin{align*}
A_4 & = \bigg\{ \max_{i} \bigg|\frac{1}{\dot{f}(\hat{u}_i)}-\frac{1}{\dot{f}({u}_i)}\bigg| = O\bigg({\frac{k\log p}{\sqrt{n}}}\bigg) \bigg\},\\
A_5 &= \bigg\{ \max_{1\le j\le p} \bigg|\frac{\|v_j\|_n}{\sqrt{n}}-F_{jj}^{1/2}   \bigg|=    O\bigg(\frac{\sqrt{k\log p}}{n^{1/4}}\bigg)\bigg\},
\end{align*}
hold with probability at least $1-O(p^{-c})$ for some constant $c>0$. 
\eel
In particular, in (C1) of Lemma \ref{lem.var}, we assume that $k = o(\sqrt{n}/\log^3 p)$, so $A_5$ in Lemma \ref{events} implies Lemma \ref{lem.var} under (C1).
On the other hand, under the bounded design, we have the following lemma.
\bel \label{events.bounded}
Under the bounded design, assume (A2) (A3) and (A4) hold, $k=o(n/\log p)$, then events $A_0, A_1,A_2$ (in Lemma \ref{events}) and
\begin{align*}
A'_3 & = \bigg\{ \max_{i,j}\big| {\hat{\eta}_{ij}}-{\eta_{ij}} \big| = O\bigg(k\sqrt{\frac{\log p}{{n}}}\bigg)\bigg\} ,\\
A'_4 & = \bigg\{ \max_{i} \bigg|\frac{1}{\dot{f}(\hat{u}_i)}-\frac{1}{\dot{f}({u}_i)}\bigg| = O\bigg(k\sqrt{\frac{\log p}{{n}}}\bigg) \bigg\},\\
A'_5 &= \bigg\{ \max_{1\le j\le p} \bigg|\frac{\|v_j\|_n}{\sqrt{n}}-F_{jj}^{1/2}   \bigg|=    O\bigg(\frac{\sqrt{k}\log^{1/4} p}{n^{1/4}}\bigg)\bigg\},
\end{align*}
hold with probability at least $1-O(p^{-c})$ for some constant $c>0$. 
\eel
In (C2) of Lemma \ref{lem.var}, we assume that $k = o(\sqrt{n}/\log^{5/2} p)$, so event $A'_5$ in Lemma \ref{events.bounded} implies Lemma \ref{lem.var} under (C2). Now we proceed to prove Lemma 1.

For event $B_1$, we first show that
\beq \label{compare.mj0}
\max_j|\check{M}_j-\tilde{M}_j| = o\bigg( \frac{1}{\sqrt{\log p}} \bigg),
\eeq
holds in probability.
To see this, note that for any $j$,
\begin{align*}
|\check{M}_j-\tilde{M}_j| &\le \bigg| \frac{\langle v_j, \epsilon\rangle}{\|v_j\|_n}- \frac{\langle v_j, \epsilon\rangle}{\sqrt{nF_{jj}}}\bigg|+\bigg| \frac{\langle v_j, \epsilon\rangle}{\sqrt{nF_{jj}}}-\frac{\sum_{i=1}^n\eta_{ij}\epsilon_i/\dot{f}(u_i)}{\sqrt{nF_{jj}}} \bigg| \\
&= T_1+T_2.
\end{align*}
It follows that
\beq \label{ineq1.1}
T_1 \le \bigg|\frac{1}{\sqrt{n}}\sum_{i=1}^n v_{ij}\epsilon_i  \bigg|\cdot \bigg| \frac{\sqrt{n}}{\|v_j\|_n}- \frac{1}{\sqrt{F_{jj}}} \bigg|.
\eeq
To bound $T_1$, by Lemma \ref{lem.var}, we only need to obtain an upper bound of $\big|\frac{1}{\sqrt{n}}\sum_{i=1}^n v_{ij}\epsilon_i  \big|$.
Note that conditional on $X$ and $\hat{\beta}$, $v_{ij}$ is fixed and $v_{ij}\epsilon_i$ are conditional independent sub-gaussian random variables. In particular, we have $\E [v_{ij}\epsilon_i|X,\hat{\beta}] =0$ and $\E[v_{ij}^2\epsilon_i^2|X,\hat{\beta}] \le v_{ij}^2$. Thus, by concentration of independent sub-gaussian random variables, for any $t\ge0$
\[
P_\theta\bigg( \frac{1}{{n}}\sum_{i=1}^n v_{ij}\epsilon_i \ge t  \bigg|X,\hat{\beta}\bigg) \le \exp\bigg( -\frac{t^2n^2}{2\sum_{i=1}^nv_{ij}^2} \bigg).
\]
It then follows that
\[
P_\theta\bigg( \frac{1}{{n}}\sum_{i=1}^n v_{ij}\epsilon_i \ge t \bigg) = \int P_\theta\bigg( \frac{1}{{n}}\sum_{i=1}^n v_{ij}\epsilon_i \ge t  \bigg|X,\hat{\beta}\bigg) dP_{X,\hat{\beta}} \le \E \exp\bigg( -\frac{t^2n^2}{2\sum_{i=1}^nv_{ij}^2} \bigg).
\]
Let $t=C\sqrt{\log p/n}$, we have
\beq \label{con.001}
P_\theta\bigg( \frac{1}{{n}}\sum_{i=1}^n v_{ij}\epsilon_i \ge C\sqrt{\frac{\log p}{n}} \bigg)  \le \E \exp\bigg( -\frac{c\log p}{2\sum_{i=1}^nv_{ij}^2/n} \bigg).
\eeq
Now under either (C1) or (C2), we have
\[
\bigg|\frac{1}{n}\sum_{i=1}^nv_{ij}^2 - \frac{1}{n}\sum_{i=1}^n{\eta}_{ij}^2/\dot{f}^2({u}_i)\bigg|\le \max_i| \hat{\eta}_{ij}^2/\dot{f}(\hat{u}_i)^2-{\eta}_{ij}^2/\dot{f}^2({u}_i) |= o_P(1).
\]
To see this, by Lemma \ref{events} and Lemma \ref{events.bounded}, we have
\begin{align*}
\max_i| \hat{\eta}_{ij}^2/\dot{f}^2(\hat{u}_i)-{\eta}_{ij}^2/\dot{f}^2({u}_i) | &\le \max_i\frac{|{\eta}_{ij}^2\dot{f}^2(\hat{u}_i)-\hat{\eta}_{ij}^2\dot{f}^2({u}_i)|}{r^2(r^2-o(1))}\\
&\le \max_i \frac{{\eta}_{ij}^2|\dot{f}^2(\hat{u}_i)-\dot{f}^2(\hat{u}_i)|+\dot{f}^2({u}_i)|\hat{\eta}_{ij}^2-{\eta}_{ij}^2|}{r^2(r^2-o(1))}\\
&= \left\{ \begin{array}{ll}
O(k\log^2p/\sqrt{n}) & \textrm{under (C1)}\\
O(k\log^{1/2}p/\sqrt{n}) & \textrm{under (C2)}
\end{array} \right.
\end{align*}
with probability at least $1-O(p^{-c})$.
By concentration inequality for sub-exponential random variables $\eta^2_{ij}/\dot{f}^2(u_i)$ (see the arguments following (\ref{subexp.conc}) in the proof of Lemma 10 for more details), we have 
\[
P_\theta\bigg(\frac{1}{n}\sum_{i=1}^n\eta_{ij}^2/\dot{f}^2(u_i) > C+\sqrt{\frac{\log p}{n}}\bigg) = O(p^{-c})
\]
for some $C,c>0$. Thus it follows that
\[
P_\theta\bigg(\frac{1}{n}\sum_{i=1}^nv_{ij}^2 > C\bigg) = O(p^{-c}).
\]
for some $C,c>0$. Now notice that
\begin{align*}
\E \exp\bigg( -\frac{c\log p}{2\sum_{i=1}^nv_{ij}^2/n} \bigg) &\le \E\bigg[\exp\bigg( -\frac{c\log p}{2\sum_{i=1}^nv_{ij}^2/n} \bigg) 1\bigg\{ \frac{1}{n}\sum_{i=1}^nv_{ij}^2 \le C \bigg\}\bigg]\\
&\quad+ \E\bigg[\exp\bigg( -\frac{c\log p}{2\sum_{i=1}^nv_{ij}^2/n} \bigg)1\bigg\{ \frac{1}{n}\sum_{i=1}^nv_{ij}^2 > C\bigg\} \bigg]\\
&\le  p^{-1/2C}+O(p^{-c'})\\
&=O(p^{-c}),
\end{align*}
by (\ref{con.001}), we have
\beq \label{con.ve}
P_\theta\bigg( \frac{1}{\sqrt{n}}\sum_{i=1}^n v_{ij}\epsilon_i \ge C\sqrt{{\log p}} \bigg) =O(p^{-c}).
\eeq
Thus, combining with Lemma \ref{lem.var}, we have
\[
T_1 \le C \sqrt{\log p}\cdot o\bigg( \frac{1}{\log p} \bigg) = o\bigg(  \frac{1}{\sqrt{\log p}} \bigg),
\]
with probability at least $1-O(p^{-c})$.
On the other hand,
\begin{align*}
T_2&\le F_{jj}^{-1/2}\bigg|\frac{1}{\sqrt{n}}\sum_{i=1}^n v_{ij}\epsilon_i  -\frac{1}{\sqrt{n}}\sum_{i=1}^n {\eta}_{ij}\epsilon_i/\dot{f}({u}_i)  \bigg|\\
&= F_{jj}^{-1/2}\bigg| \frac{1}{\sqrt{n}} \sum_{i=1}^n\epsilon_i\bigg[ \frac{\hat{\eta}_{ij}}{\dot{f}(\hat{u}_i)}- \frac{{\eta}_{ij}}{\dot{f}({u}_i)} \bigg] \bigg|.
\end{align*}
Following the same conditional argument as (\ref{con.001}), we have
\[
P_\theta\bigg( \frac{1}{\sqrt{n}} \sum_{i=1}^n\epsilon_i\bigg[ \frac{\hat{\eta}_{ij}}{\dot{f}(\hat{u}_i)}- \frac{{\eta}_{ij}}{\dot{f}({u}_i)} \bigg] \ge t\bigg) \le \E \exp\bigg( -\frac{t^2}{2\sum_{i=1}^n \alpha^2_{ij}/n} \bigg)
\]
where $\alpha_{ij}= \frac{\hat{\eta}_{ij}}{\dot{f}(\hat{u}_i)}- \frac{{\eta}_{ij}}{\dot{f}({u}_i)}$. Under (C2), we have $\alpha_{ij}^2 =O\big(\frac{ k^2\log p}{n}\big).$
Then 
\[
P_\theta\bigg( \frac{1}{\sqrt{n}} \sum_{i=1}^n\epsilon_i\bigg[ \frac{\hat{\eta}_{ij}}{\dot{f}(\hat{u}_i)}- \frac{{\eta}_{ij}}{\dot{f}({u}_i)} \bigg] \ge t\bigg) \le  \exp\bigg( -\frac{nt^2}{2k^2\log p} \bigg)+O(p^{-c}).
\]
Let $t=k\log p/\sqrt{n}$, we have
\[
P_\theta\bigg( \frac{1}{\sqrt{n}} \sum_{i=1}^n\epsilon_i\bigg[ \frac{\hat{\eta}_{ij}}{\dot{f}(\hat{u}_i)}- \frac{{\eta}_{ij}}{\dot{f}({u}_i)} \bigg] \ge \frac{k\log p}{\sqrt{n}}\bigg) =O(p^{-c}).
\]
Therefore $T_2 = O\big( \frac{k\log p}{\sqrt{n}} \big) = o(1/\sqrt{\log p})$ with probability at least $1-O(p^{-c})$ as long as $k=o(\sqrt{n}/\log^{3/2}p)$. Under (C1), similar argument yields $T_2=o(1/\sqrt{\log p})$ with probability at least $1-O(p^{-c})$ as long as $k=o(\sqrt{n}/\log^2p)$. Using a union bound argument across $j=1,...,p$, we prove that (\ref{compare.mj0}) holds in probability. Using the same argument, we can prove 
\beq \label{m.check.hp}
P_\theta\bigg(\max_j|\check{M}_j| >C\sqrt{\log p}\bigg) =O(p^{-c}).
\eeq
Therefore, we have
\begin{align*}
|\check{M}_n-\tilde{M}_n| \le \max_j | \check{M}_j^2-\tilde{M}_j^2| \le C(\max_j |\tilde{M}_j | )\cdot \max_j|\check{M}_j-\tilde{M}_j| =o(1)
\end{align*}
with probability at least $1-O(p^{-c})$. This completes the proof of event $B_1$.

For event $B_2$, note that
\beq \label{m0-m.0}
|\tilde{M}_n-M_n| \le \max_j | \tilde{M}_j^2-M_j^2| \le C(\max_j |\check{M}_j|)\cdot \max_j\bigg( \frac{|\langle v_j,Re_i\rangle|}{\|v_j\|_n}+\frac{|\langle v_j, \bold{h}_{-j}\rangle|}{\|v_j\|_n} \bigg).
\eeq
To bound $\max_j {|\langle v_j, Re\rangle|}/{\|v_j\|_n}$, by Lemma \ref{lem.var} and mean value theorem,
\begin{align*}
\frac{|\langle v_j, Re\rangle|}{\|v_j\|_n}  &\le \frac{\big|\sum_{i=1}^n v_{ij}(\dot{f}(\hat{u}_i)-\dot{f}(u_i^*))(\hat{u}_i-u_i)\big|}{\sqrt{n}(F_{jj}^{1/2}-o_P(1))}
\end{align*}
Under (C1), $\max_{i,j}|v_{ij}|=O_P(\sqrt{\log p})$ and thereby
\begin{align*}
&\bigg|\sum_{i=1}^n v_{ij}(\dot{f}(\hat{u}_i)-\dot{f}(u_i^*))(\hat{u}_i-u_i)\bigg| \le \sum_{i=1}^n(\hat{u}_i-u_i)^2 \cdot \max_{i,j}|v_{ij}|=\|X(\hat{\beta}-\beta)\|_2^2\cdot O(\sqrt{\log p})\\
& = O(k\log^{3/2}p)
\end{align*}
with probability at least $1-O(p^{-c})$. Thus 
\[
\max_j\frac{|\langle v_j, Re\rangle|}{\|v_j\|_n} = O\bigg( \frac{k\log^{3/2}p}{\sqrt{n}} \bigg)
\]
in probability. 
Under (C2), $\max_{i,j}|v_{ij}|=O_P(1)$ and thereby
\begin{align*}
&\bigg|\sum_{i=1}^n v_{ij}(\dot{f}(\hat{u}_i)-\dot{f}(u_i^*))(\hat{u}_i-u_i)\bigg| \le \sum_{i=1}^n(\hat{u}_i-u_i)^2 \cdot \max_{i,j}|v_{ij}|=\|X(\hat{\beta}-\beta)\|_2^2\cdot O(1)\\
& = O(k\log p)
\end{align*}
with probability at least $1-O(p^{-c})$. Thus 
\beq 
\max_j\frac{|\langle v_j, Re\rangle|}{\|v_j\|_n} = O\bigg( \frac{k\log p}{\sqrt{n}} \bigg)
\eeq
In general, either (C1) or (C2) implies that 
\beq \label{reminder}
\max_j {|\langle v_j, Re\rangle|}/{\|v_j\|_n} =o(\log^{-3/2} p)
\eeq
with probability at least $1-O(p^{-c})$.
On the other hand, to bound $\max_j{|\langle v_j, \bold{h}_{-j}\rangle|}/{\|v_j\|_n}$, by Proposition 1 (ii) in \cite{zhang2014confidence}, we know that if we choose $\lambda=C\sqrt{{\log p}/{n}}$, then under (C1) or (C2)
\beq \label{zhangzhang}
\max_{k\ne j}\frac{\langle \hat{\eta}_j, \bx_k\rangle }{\|\hat{\eta}_j\|_2} \le C_1\sqrt{2\log p},\quad\quad \frac{\|\hat{\eta}_j\|_2}{\langle \hat{\eta}_j, \bx_j \rangle} \le \frac{C_2}{\sqrt{n}}
\eeq
with probability at least $1-O(p^{-c})$.
Note that
\[
\|\hat{\eta}_j\|_2 = \sqrt{\sum_{i=1}^n \hat{\eta}_{ij}^2} = \sqrt{\sum_{i=1}^n \dot{f}^2(\hat{u}_i)v_{ij}^2} \le  \sqrt{\sum_{i=1}^n \dot{f}(\hat{u}_i)v_{ij}^2} = \|v_j\|_n,
\]
we have
\beq \label{eta.bnd}
\eta_j = \max_{k\ne j} \frac{\langle v_j, \bx_k\rangle_n }{\|v_j\|_n} \le C_1\sqrt{2\log p}
\eeq
in probability.
Therefore under either (C1) or (C2)
\begin{align} \label{reminder2}
&\frac{|\langle v_j, \bold{h}_{-j}\rangle|}{\|v_j\|_n}  \le \|v_j\|_n^{-1}\bigg|\sum_{i=1}^nv_{ij}\dot{f}(\hat{u}_i)X_{i,-j}^\top (\hat{\beta}_{-j}-\beta_{-j}) \bigg|\le  \max_{k\ne j}\frac{|\langle v_j,\bold{x}_k \rangle_n|}{\|v_j\|_n}\cdot \|\hat{\beta}-\beta\|_1 \nonumber\\
&=\eta_j\|\hat{\beta}-\beta\|_1=O\bigg( \frac{k\log p}{\sqrt{n}}\bigg)
\end{align}
with probability at least $1-O(p^{-c})$.
Back to (\ref{m0-m.0}), note that $\max_j |\tilde{M}_n| \le \max_j|\check{M}_n| +o_P(1) =O_P(\sqrt{\log p})$, we have
\[
|\tilde{M}_n-M_n| = o\bigg( \frac{1}{{\log p}}\bigg)
\]
with probability at least $1-O(p^{-c})$.
\qed


\paragraph{Proof of Lemma 2.} The lemma is proved under the Gaussian design. For the bounded design, by definition $\hat{M}_j$ is essentially the same as $\check{M}_j$.
Note that
\begin{align*} 
\max_{1\le j\le p} \frac{1}{\sqrt{n}} \sum_{i=1}^n \E [|v^0_{ij}\epsilon_i| 1\{ |v^0_{ij}\epsilon_i| \ge \tau_n\}] &\le Cn^{1/2}\max_{i,j}\E [|v^0_{ij}\epsilon_i| 1\{ |v^0_{ij}\epsilon_i| \ge \tau_n\}] \\
&\le Cn^{1/2}(p+n)^{-1}\max_{i,j}\E [|v^0_{ij}\epsilon_i| e^{|v^0_{ij}\epsilon_i|}]\\
&\le Cn^{1/2}(p+n)^{-1},
\end{align*}
where the last inequality follows from
\[
\E [|v^0_{ij}\epsilon_i| e^{|v^0_{ij}\epsilon_i|}] \le C_1 \sqrt{\E (v^0_{ij})^2}\sqrt{\E \exp(2|v^0_{ij}\epsilon_i|)} \le C_2 
\]
by sub-gaussianity of $v^0_{ij}$. Hence, if $ \max_{i,j} |v^0_{ij}\epsilon_i| \le \tau_n$, then
\[
\hat{Z}_{ij} = v_{ij}^0\epsilon_i-\E[v_{ij}^0\epsilon_i1\{|v_{ij}^0\epsilon_i|\le \tau_n\}]
\]
and thereby
\begin{align*}
\max_j |\hat{M}_j-\check{M}_j| &\le \max_{1\le j\le p} \bigg|\frac{1}{\sqrt{nF_{jj}}} \sum_{i=1}^n \E [v^0_{ij}\epsilon_i 1\{ |v^0_{ij}\epsilon_i| \le \tau_n\}] \bigg| \\
& =\max_{1\le j\le p} \bigg|\frac{1}{\sqrt{nF_{jj}}} \sum_{i=1}^n \E [v^0_{ij}\epsilon_i 1\{ |v^0_{ij}\epsilon_i| \ge \tau_n\}] \bigg| \\
&\le \max_{1\le j\le p} \frac{1}{\sqrt{nF_{jj}}} \sum_{i=1}^n \E [|v^0_{ij}\epsilon_i| 1\{ |v^0_{ij}\epsilon_i| \ge \tau_n\}]\\
&\le Cn^{1/2}(p+n)^{-1}\\
&= O(1/\log p).
\end{align*}
Then 
we have
\beq \label{hatcheck}
P_\theta\bigg(\max_j |\hat{M}_j-\check{M}_j| \ge C(\log p)^{-1}  \bigg) \le P\bigg( \max_{i,j} |v_{ij}\epsilon_i| \ge \tau_n \bigg) = O(p^{-c}).
\eeq
Now by the fact that
\[
|\hat{M}_n-\check{M}_n| \le 2\max_j|\check{M}_i|\max_{j}|\check{M}_j-\hat{M}_j| + \max_j |\check{M}_j-\hat{M}_j|^2,
\]
it suffices to apply (\ref{hatcheck}) and (\ref{m.check.hp}) in the proof of Lemma 1.
\qed

\paragraph{Proof of Lemma 6.} By definition, we have
\begin{align}
\chi^2(g,f) &= \int\frac{g^2}{f}-1 \nonumber \\
&= \frac{1}{{p \choose k}^{2}}\int \frac{(\sum_{\beta\in \mathcal{H}_1} \prod_{i=1}^n p(X_i,y_i;\beta))^2}{\prod_{i=1}^np(X_i,y_i)} -1 \nonumber \\
&= \frac{1}{{p \choose k}^{2}} \sum_{\beta\in \mathcal{H}_1} \sum_{\beta' \in \mathcal{H}_1}  \prod_{i=1}^n \int \frac{ p(X_i,y_i;\beta)p(X_i,y_i;\beta')}{p(X_i,y_i)} -1.
\end{align}
Note that 
\begin{align}
&\int \frac{ p(X_i,y_i;\beta)p(X_i,y_i;\beta')}{p(X_i,y_i)} \\
&=\frac{1}{(2\pi)^{p/2}} \int\int \frac{2\exp(-\frac{1}{2}X_i^\top X_i+y_iX_i^\top (\beta+\beta'))}{[1+\exp(X_i^\top \beta)][1+\exp(X_i^\top \beta')]}dy_idX_i\nonumber  \\
& =\frac{1}{(2\pi)^{p/2}}  \int \frac{2\exp(-\frac{1}{2}X_i^\top X_i+X_i^\top (\beta+\beta'))}{[1+\exp(X_i^\top \beta)][1+\exp(X_i^\top \beta')]}dX_i\nonumber \\
&\quad + \frac{1}{(2\pi)^{p/2}} \int \frac{2\exp(-\frac{1}{2}X_i^\top X_i)}{[1+\exp(X_i^\top \beta)][1+\exp(X_i^\top \beta')]}dX_i \nonumber \\
& =  \E h(X;\beta,\beta')
\end{align}
where in the last equality, the expectation is with respect to a standard multivariate normal random vector $X \sim N(0, \bold{I}_p)$ and
\begin{align*}
h(X;\beta,\beta') &=: \frac{2(1+e^{X^\top (\beta+\beta')})}{(1+e^{X^\top \beta})(1+e^{X^\top \beta'})} =  1+\frac{e^{X^\top \beta}-1}{e^{X^\top \beta}+1}\frac{e^{X^\top \beta'}-1}{e^{X^\top \beta'}+1}\\
&= 1+\tanh\bigg(\frac{{X^\top \beta}}{2}\bigg)\tanh\bigg(\frac{{X^\top \beta'}}{2}\bigg)
\end{align*}

\bel \label{tansh}
If $(X,Y)\sim N(0, \Sig)$ with $\Sig=\sigma^2\begin{pmatrix}
1& \rho\\
\rho& 1
\end{pmatrix}$ for some $\sigma^2\le 1$, then it follows
\[
\E \tanh\bigg(\frac{X}{2}\bigg)\tanh\bigg(\frac{Y}{2}\bigg) \le 6\sigma^2\rho.
\]
\eel
Now since $X_i^\top \beta \sim N(0,k\rho^2)$, where we can choose $\rho$ such that $k\rho^2 \le 1$. By Lemma \ref{tansh}, let $j = |\text{supp}(\beta)\cap\text{supp}(\beta')| = |I\cap I'|$ be the number of intersected components between $\beta$ and $\beta'$, we have
\begin{align*}
\chi^2(g,f)& \le \frac{1}{{p \choose k}^{2}}\sum_{\beta\in \mathcal{H}_1} \sum_{\beta' \in \mathcal{H}_1} \bigg( 1+6\beta^\top \beta' \bigg)^n-1 = \frac{1}{{p \choose k}^{2}}\sum_{\beta\in \mathcal{H}_1} \sum_{\beta' \in \mathcal{H}_1}  \bigg( 1+6j\rho^2 \bigg)^n-1
\end{align*}
Note that for $\beta,\beta'$ uniformly picked from $\mathcal{H}_1$, $j$ follows a hypergeometric distribution 
\[
P(J=j) = \frac{{k \choose j}{p-k \choose k-j}}{{p \choose k}},\quad j=0,1,...,k.
\]
Then
\begin{align*}
\chi^2(g,f)&\le \E ( 1+ 6\rho^2J)^n-1= \E\exp(n\log (1+ 6\rho^2J))-1\le \E e^{6n\rho^2J}-1.
\end{align*}
As shown on page 173 of \citep{aldous1985exchangeability}, $J$ has the same distribution as the random variable $\E(Z|\mathcal{B}_n)$ where $Z$ is a binomial random variable of parameters $(k,k/p)$ and $\mathcal{B}_n$ some suitable $\sigma$-algebra. Thus by Jensen's inequality we have 
\begin{align*}
\E e^{6nJ\rho^2} &\le \bigg( 1-\frac{k}{p}+\frac{k}{p}e^{6n\rho^2} \bigg)^{k}.
\end{align*}
Let
\[
\rho^2 = \frac{1}{6n}\log \bigg(1+\frac{p}{h(\eta)k^2} \bigg),
\]
where $h(\eta)= [\log(\eta^2+1)]^{-1}$ and $\eta=1-\alpha-\delta$, 
we have
\[
\E e^{6n\rho^2J} \le e^{1/h(\eta)}, 
\]
so that 
\[
\chi^2(g,f) \le \eta^2 = (1-\alpha-\delta)^2.
\]
\qed

\paragraph{Proof of Lemma \ref{thm3.lem2}.}
Note that
\[
|M'_j| \le  \frac{|\langle v_j,\bold{\epsilon}\rangle|}{\|v_j\|_n}+\frac{|\langle v_j, Re \rangle|}{\|v_j\|_n} + \frac{|\langle v_j,\bold{h}_{-j}\rangle_n|}{\|v_j\|_n}.
\]
We bound the above three terms one by one. Firstly, by concentration of sub-exponential random variables $\eta_{ij}^2$ (see (\ref{subexp.conc}) in the proof of Lemma 10 for details) and (2.18), we have 
\beq \label{etahat.bnd}
P_\theta\bigg(\bigg|\frac{1}{n}\sum_{i=1}^n\hat{\eta}_{ij}^2-\E \eta_{ij}^2\bigg| \ge (\log p)^{-1} \bigg)=O(p^{-c})
\eeq
Then we have
\[
\frac{|\langle v_j,\bold{\epsilon}\rangle|}{\|v_j\|_n}\le \frac{n^{-1/2}\sum_{i=1}^n \hat{\eta}_{ij}\epsilon_i/\dot{f}(\hat{u_i})}{\sqrt{\sum_{i=1}^n\hat{\eta}_{ij}^2/n}}\le \frac{C}{\sqrt{n}}\sum_{i=1}^n \hat{\eta}_{ij}\epsilon_i/\dot{f}(\hat{u}_i)\equiv\frac{C}{\sqrt{n}}\sum_{i=1}^n\xi_i.
\]
Conditional on $X$ and $\hat{\beta}$, we have $\E[\xi_i|X,\hat{\beta}]=0$ and $\E[\xi_i^2|X,\hat{\beta}] \le \hat{\eta}_{ij}^2/\dot{f}^2(\hat{u}_i)=\alpha_{ij}(n)$. By concentration inequality for independent sub-gaussian random variables $\xi_i|X,\hat{\beta}$, we have  for any $t\ge0$
\[
P_\theta\bigg( \frac{1}{{n}}\sum_{i=1}^n \xi_i \ge t  \bigg|X,\hat{\beta}\bigg) \le \exp\bigg( -\frac{t^2n^2}{2\sum_{i=1}^n\alpha_{ij}(n)} \bigg).
\]
It then follows that
\[
P_\theta\bigg( \frac{1}{{n}}\sum_{i=1}^n \xi_i \ge t \bigg) = \int P_\theta\bigg( \frac{1}{{n}}\sum_{i=1}^n \xi_i \ge t  \bigg|X,\hat{\beta}\bigg) dP_{X,\hat{\beta}} \le \E \exp\bigg( -\frac{t^2n^2}{2\sum_{i=1}^n\alpha_{ij}(n)} \bigg).
\]
Let $t=C\sqrt{\log p/n}$, we have
\beq \label{con.001}
P_\theta\bigg( \frac{1}{{n}}\sum_{i=1}^n \xi_i \ge C\sqrt{\frac{\log p}{n}} \bigg)  \le \E \exp\bigg( -\frac{c\log p}{2\sum_{i=1}^n\alpha_{ij}(n)/n} \bigg).
\eeq
Now since with probability at least $1-O(n^{-c})$, $\alpha_{ij}(n) \le \hat{\eta}_{ij}^2 L(n)^{-2}$, or
\[
P_\theta\bigg(\frac{1}{n}\sum_{i=1}^n\alpha_{ij}(n) \ge L(n)^{-2}\frac{1}{n}\sum_{i=1}^n\hat{\eta}_{ij}^2 \bigg) = O(n^{-c}),
\]
by (\ref{etahat.bnd}), we have
\[
P\bigg(\frac{1}{n}\sum_{i=1}^n\alpha_{ij}(n) \ge CL(n)^{-2}\bigg) = O(n^{-c})
\]
for some $C,c>0$.
Now notice that
\begin{align*}
\E \exp\bigg( -\frac{c\log p}{2\sum_{i=1}^nv_{ij}^2/n} \bigg) &\le \E\bigg[\exp\bigg( -\frac{c\log p}{2\sum_{i=1}^n\alpha_{ij}/n} \bigg) 1\bigg\{ \frac{1}{n}\sum_{i=1}^n\alpha_{ij} \le C L(n)^{-2} \bigg\}\bigg]\\
&\quad+ \E\bigg[\exp\bigg( -\frac{c\log p}{2\sum_{i=1}^n\alpha_{ij}/n} \bigg)1\bigg\{ \frac{1}{n}\sum_{i=1}^n\alpha_{ij} > CL(n)^{-2}\bigg\} \bigg]\\
&\le  p^{-1/(2CL^{-2}(n))}+O(n^{-c'})\\
&=O(n^{-c}),
\end{align*}
where we used the fact that
\[
p^{-1/(2CL^{-2}(n))} \asymp p^{-\exp(-c_1\|\beta\|_2\sqrt{\log n})} \lesssim n^{-c}
\]
for sufficiently small $c>0$, as long as $\|\beta\|_2 =O\big(\frac{\log\log p}{\sqrt{\log n}}\big)$ and $\log p \gtrsim \log^{1+\delta} n$. 
As a result, by (\ref{con.001}), we have
\beq \label{con.ve1}
P_\theta\bigg( \frac{1}{\sqrt{n}}\sum_{i=1}^n \xi_i \ge C\sqrt{{\log p}} \bigg) =O(n^{-c}).
\eeq
To bound ${|\langle v_j, Re\rangle|}/{\|v_j\|_n}$, by mean value theorem,
\begin{align*}
\frac{|\langle v_j, Re\rangle|}{\|v_j\|_n}  &\le \frac{n^{-1/2}\big|\sum_{i=1}^n v_{ij}(\dot{f}(\hat{u}_i)-\dot{f}(u_i^*))(\hat{u}_i-u_i)\big|}{\sqrt{\sum_{i=1}^n\hat{\eta}_{ij}^2/n}}
\end{align*}
Note that $\max_{i}|v_{ij}|=O(\sqrt{\log p}L^{-1}(n))$ with probability at least $1-O(n^{-c})$, thereby 
\begin{align*}
&\bigg|\sum_{i=1}^n v_{ij}(\dot{f}(\hat{u}_i)-\dot{f}(u_i^*))(\hat{u}_i-u_i)\bigg| \le \sum_{i=1}^n(\hat{u}_i-u_i)^2 \cdot \max_{i,j}|v_{ij}|\\
&=\|X(\hat{\beta}-\beta)\|_2^2\cdot O(L^{-1}(n)\sqrt{\log p})= O(k\log^{3/2}pL^{-1}(n))
\end{align*}
with probability at least $1-O(n^{-c})$. Since $\|\beta\|_2\le C\big(\frac{\log\log p}{\sqrt{\log n}}\big)$, for some $C\le\sqrt{2/\lambda_{\max}(\Sig)}$, we have 
\beq \label{2}
\frac{|\langle v_j, Re\rangle|}{\|v_j\|_n} = O\bigg( \frac{k\log^{3/2}p}{L(n)\sqrt{n}} \bigg)=o(\sqrt{\log p})
\eeq
with probability at least $1-O(n^{-c})$.
Finally, to bound $\max_j{|\langle v_j, \bold{h}_{-j}\rangle|}/{\|v_j\|_n}$, by (\ref{eta.bnd}) we have
\begin{align} \label{3}
&\frac{|\langle v_j, \bold{h}_{-j}\rangle|}{\|v_j\|_n} =O\bigg( \frac{k\log p}{\sqrt{n}}\bigg)=o(1)
\end{align}
with probability at least $1-O(n^{-c})$. Combining (\ref{con.ve1}) (\ref{2}) and (\ref{3}), we have proven Lemma \ref{thm3.lem2}.
\qed

\paragraph{Proof of Lemma \ref{events}.} Event $A_0$ and $A_2$ follows from Corollary 1, 2 and 5 of \cite{negahban2010unified}. For the event $A_1$, by Theorem 1.6 of \cite{zhou2009restricted}, the condition of Lemma \ref{events} implies the restricted eigenvalue condition, which, by Lemma 2.1 and Figure 1 of \cite{van2009conditions}, implies event $A_1$. For event $A_3$, note that under $A_2$ we have
\[
\max_{i,j}|\hat{\eta}_{ij} -\eta_{ij}|=\max_{i,j} |X_{i,-j}(\hat{\gamma}_j-\gamma_j)| \le \max_{i,j}\|X_{i,-j}\|_\infty\max_{j} \|\hat{\gamma}_j-\gamma_j\|_1 \le Ck\frac{\log p}{\sqrt{n}}
\]
where the last inequality follows from that fact that
\beq \label{gaussian.tail}
P_\theta\bigg( \max_{1\le i\le p}X_i \ge \sqrt{C\log p}  \bigg) \le \frac{1}{p^c}
\eeq
for some sufficiently large constant $C,c>0$, which is a consequence of
the Gaussian tail probability bound $1-\Phi(x) \le \frac{1}{x}\phi(x)$ by taking $x=\sqrt{C \log p}$ for some sufficiently large $C>0$.

For event $A_4$, since $\|X\beta\|_\infty \le c_2$ for some constant $c_2>0$, there exists some constant $0<\kappa<1$ such that $\kappa<|f(u_i)|<1-\kappa$ and thereby $\dot{f}(u_i)\ge \kappa(1-\kappa)$ for all $i$. $A_4$ then follows from the following lemma, event $A_0$ and (\ref{gaussian.tail}).
\bel \label{lip.lem}
Let $f(x) = \frac{e^{x}}{1+e^x}$, then uniformly over $a, b\in \R$, it holds that
\beq \label{lip.cond}
\bigg| \frac{1}{\dot{f}(a)}-\frac{1}{\dot{f}(b)}\bigg| \le \frac{\max\{\dot{f}(a), \dot{f}(b)\}}{\dot{f}(a) \dot{f}(b)}|a-b|\le \frac{1}{\dot{f}(a) \dot{f}(b)}|a-b|.
\eeq
\eel
For event $A_5$, by the fact that $v_{ij}=\hat{\eta}_{ij}/\dot{f}(\hat{u}_i)$, it follows that
\begin{align*}
&\bigg|\frac{\|v_j\|_n}{\sqrt{n}}-F_{jj}^{1/2}   \bigg|=\bigg|\bigg[{\frac{1}{n}\sum_{i=1}^nv_{ij}^2\dot{f}(\hat{u}_i)}\bigg]^{1/2}-F_{jj}^{1/2}   \bigg| \le \bigg|\frac{1}{n}\sum_{i=1}^nv_{ij}^2\dot{f}(\hat{u}_i)-F_{jj}   \bigg|^{1/2}\\
&\le \bigg|\frac{1}{n}\sum_{i=1}^n\hat{\eta}_{ij}^2/\dot{f}(\hat{u}_i)-\frac{1}{n}\sum_{i=1}^n\hat{\eta}_{ij}^2/\dot{f}({u}_i)  \bigg|^{1/2}+\bigg|\frac{1}{n}\sum_{i=1}^n\hat{\eta}_{ij}^2/\dot{f}({u}_i)-F_{jj}   \bigg|^{1/2}\\
&\le  \bigg|\frac{1}{n}\sum_{i=1}^n\hat{\eta}_{ij}^2\bigg[\frac{1}{\dot{f}(\hat{u}_i)}-\frac{1}{\dot{f}({u}_i)}\bigg] \bigg|^{1/2}+\bigg|\frac{1}{n}\sum_{i=1}^n(\hat{\eta}_{ij}^2-\eta_{ij}^2)/\dot{f}({u}_i)\bigg|^{1/2}\\
&\quad+\bigg|\frac{1}{n}\sum_{i=1}^n{\eta}_{ij}^2/\dot{f}({u}_i)-F_{jj}   \bigg|^{1/2}\\
&=I_1+I_2+I_3.
\end{align*}
To bound $I_2$, note that $\|X\beta\|_\infty \le c$ implies $\max_{i}\dot{f}(u_i) \ge r$ for some constant $r>0$, and that $\hat{\eta}_j - \eta_j=X_{-j}(\hat{\gamma}_j-\gamma_j)$, we have
\begin{align} 
I_2^2 &\le \frac{1}{rn}\sum_{i=1}^n |\hat{\eta}_{ij}^2-\eta_{ij}^2|\le \frac{1}{rn}\sum_{i=1}^n[ |\hat{\eta}_{ij}-\eta_{ij}|^2+2|\hat{\eta}_{ij}-\eta_{ij}|\cdot|\eta_{ij}|]\nonumber \\
&\le \frac{1}{rn} \| X_{-j}(\hat{\gamma}-\gamma_j) \|_2^2+ \frac{2C\sqrt{\log p}}{rn} \| X_{-j}(\hat{\gamma}-\gamma_j) \|_1 \nonumber \\
&\le \frac{1}{rn} \| X_{-j}(\hat{\gamma}-\gamma_j) \|_2^2+\frac{2C\sqrt{\log p}}{r\sqrt{n}} \| X_{-j}(\hat{\gamma}-\gamma_j) \|_2.
\end{align}
Therefore, by event $A_1$, as long as $k<n$,
\beq \label{I_21}
I_2^2 \le C_1k\frac{\log p}{n}+ C_2\sqrt{k}\frac{\log p}{\sqrt{n}} = O\bigg( \sqrt{k}\frac{\log p}{\sqrt{n}}\bigg)
\eeq
with probability at least $1-O(p^{-c})$ for some $c>0$.
By $A_4$ and (\ref{I_21}), we have, with probability at least $1-O(p^{-c})$ for some $c>0$,
\[
I_1^2 \le  \frac{1}{n}\sum_{i=1}^n\hat{\eta}_{ij}^2 \cdot Ck{\frac{\log p}{\sqrt{n}}} \le \bigg[\frac{1}{n}\sum_{i=1}^n{\eta}_{ij}^2+o(1)\bigg]\cdot Ck{\frac{\log p}{\sqrt{n}}} \le C'k\frac{\log p}{\sqrt{n}},
\]
where the last inequality follows from the concentration inequality
\beq \label{subexp.conc}
P_\theta\bigg(\bigg|\frac{1}{n}\sum_{i=1}^n \eta_{ij}^2 - \E \eta_{ij}^2\bigg| \ge \sqrt{\frac{\log p}{n}} \bigg) = O(p^{-c}).
\eeq
To show this, we need to introduce the following norms for random variables. The sub-gaussian norm of a random variable $U$ is defined as $\| U\|_{\psi_2} = \sup_{q\ge 1}\frac{1}{\sqrt{q}}(\E|U|^q)^{1/q}$, and the sub-exponential norm of a random variable is defined as $\| U\|_{\psi_1} = \sup_{q\ge 1}q^{-1}(\E|U|^q)^{1/q}$. By definition $\eta_{ij}$ are sub-gaussian with $\|\eta_{ij}\|_{\psi_2} < C<\infty$ and therefore 
\[
\| \eta_{ij}^2\|_{\psi_1} =  \sup_{q\ge 1}q^{-1}(\E|\eta_{ij}|^{2q})^{1/q} =  \sup_{q\ge 1}[q^{-1/2}(\E|\eta_{ij}|^{2q})^{1/2q}]^2 = \| \eta_{ij}\|_{\psi_2}^{2}<C^2<\infty.
\]
So $\eta_{ij}^2$ with $i=1,...,n$ are i.i.d. sub-exponential random variables. Then (\ref{subexp.conc}) follows from standard concentration inequality for sub-exponential random variables (see, for example, Proposition 5.16 in \cite{vershynin2010introduction}).
Similarly, we can show $\eta_{ij}^2/\dot{f}(u_i)$ are sub-exponential and therefore
\begin{align*}
I_3^2 = \bigg|\frac{1}{n}\sum_{i=1}^n{\eta}_{ij}^2/\dot{f}({u}_i)-F_{jj}   \bigg| = O\bigg( \sqrt{\frac{\log p}{n}} \bigg)
\end{align*}
with probability at least $1-O(p^{-c})$ for some $c>0$. Thus, $I_1+I_2+I_3 = O\big( \frac{\sqrt{k\log p}}{n^{1/4}}\big) $.
\qed

\paragraph{Proof of Lemma \ref{events.bounded}.} Events $A_0$ $A_1$ and $A_2$ follow the same argument as in Lemma \ref{events}. For event $A'_3$, by $A_1, A_2$ and boundedness of $X$, we have
\[
\max_{i,j}|\hat{\eta}_{ij} -\eta_{ij}|=\max_{i,j} |X_{i,-j}(\hat{\gamma}_j-\gamma_j)| \le \max_{i,j}\|X_{i,-j}\|_\infty\max_{j} \|\hat{\gamma}_j-\gamma_j\|_1 \le Ck\sqrt{\frac{\log p}{{n}}}
\]
For event $A'_4$, by (A4), there exists some constant $r>0$ such that $\dot{f}(u_i)\ge r$ for all $i$ with probability at least $1-O(p^{-c})$. $A'_4$ then follows from Lemma \ref{lip.lem}.
For event $A'_5$, as the proof of $A_5$ in Lemma \ref{events}, we have
\begin{align*}
\bigg|\sqrt{\frac{1}{n}\sum_{i=1}^nv_{ij}^2\dot{f}(\hat{u}_i)}-F_{jj}^{1/2}   \bigg| &\le  \bigg|\frac{1}{n}\sum_{i=1}^n\hat{\eta}_{ij}^2\bigg[\frac{1}{\dot{f}(\hat{u}_i)}-\frac{1}{\dot{f}({u}_i)}\bigg] \bigg|^{1/2}+\bigg|\frac{1}{n}\sum_{i=1}^n(\hat{\eta}_{ij}^2-\eta_{ij}^2)/\dot{f}({u}_i)\bigg|^{1/2}\\
&\quad+\bigg|\frac{1}{n}\sum_{i=1}^n{\eta}_{ij}^2/\dot{f}({u}_i)-F_{jj}   \bigg|^{1/2}\\
&=I_1+I_2+I_3.
\end{align*}
To bound $I_2$, note that $P(\max_{i}\dot{f}(u_i)\le r)=O(p^{-c})$ , and that $\hat{\eta}_j - \eta_j=X_{-j}(\hat{\gamma}_j-\gamma_j)$, by $A_1$,
\begin{align}
I_2^2 &\le \frac{1}{rn}\sum_{i=1}^n |\hat{\eta}_{ij}^2-\eta_{ij}^2|\le \frac{1}{rn}\sum_{i=1}^n[ |\hat{\eta}_{ij}-\eta_{ij}|^2+2|\hat{\eta}_{ij}-\eta_{ij}|\cdot|\eta_{ij}|]\nonumber \\
&\le \frac{1}{rn} \| X_{-j}(\hat{\gamma}-\gamma_j) \|_2^2+ \frac{2C}{rn} \| X_{-j}(\hat{\gamma}-\gamma_j) \|_1 \nonumber \\
&\le \frac{1}{rn} \| X_{-j}(\hat{\gamma}-\gamma_j) \|_2^2+\frac{2C}{r\sqrt{n}} \| X_{-j}(\hat{\gamma}-\gamma_j) \|_2 \nonumber \\
&\le C_1k\frac{\log p}{n}+ C_2\sqrt{k\frac{\log p}{n}} = O\bigg( \sqrt{k\frac{\log p}{n}}\bigg)
\end{align}
with probability at least $1-O(p^{-c})$ for some $c>0$.
For $I_1$, by $A'_4$ and boundedness of $X$, we have, with probability at least $1-O(p^{-c})$ for some $c>0$,
\[
I_1^2 \le  \frac{1}{n}\sum_{i=1}^n\hat{\eta}_{ij}^2 \cdot Ck\sqrt{\frac{\log p}{{n}}} =O\bigg(k\sqrt{\frac{\log p}{{n}}}  \bigg).
\]
Finally, by concentration inequality for sub-exponential random variables $\eta_{ij}^2/\dot{f}(u_i)$ for $i=1,...,n$, we have
\begin{align*}
I_3^2 = \bigg|\frac{1}{n}\sum_{i=1}^n{\eta}_{ij}^2/\dot{f}({u}_i)-F_{jj}   \bigg| = O\bigg( \sqrt{\frac{\log p}{n}} \bigg)
\end{align*}
with probability at least $1-O(p^{-c})$ for some $c>0$. Thus, $I_1+I_2+I_3 = O\big( \frac{\sqrt{k}\log^{1/4} p}{n^{1/4}}\big)$.
\qed

\paragraph{Proof of Lemma \ref{tansh}.} By normalization, we only need to consider $(X,Y)$ with $\text{Var}(X)=\text{Var}(Y)=1$ and $\E XY =\rho$ and prove
\beq
\E \tanh\bigg(\frac{\sigma X}{2}\bigg)\tanh\bigg(\frac{\sigma Y}{2}\bigg) \le 10\sigma^2 \rho.
\eeq
Note that the inner product
\[
\langle X, Y \rangle = \E XY
\]
defines a Hilbert space on $L^2(\Omega, \mathcal{F},\mu)$. Then the above inequality is equivalent to
\[
\bigg\langle \tanh\bigg(\frac{\sigma X}{2}\bigg), \tanh\bigg(\frac{\sigma Y}{2}\bigg) \bigg\rangle \le \sigma^2\langle X, Y \rangle. 
\]
Consider the Hermite polynomials $H_n(x), x\in\R, n=0,1,...$ which are defined as
\[
H_n = \frac{(-1)^n}{n!}e^{x^2/2}\frac{d^n}{dx^n}(e^{-x^2/2}),
\]
so that in particular $H_0(x)=1, H_1(x)=x, H_2(x) = (x^2-1)/2$, and in general $H_n(x)$ is a polynomial of order $n$. The Hermite polynomials satisfy the following basic identities
\begin{align}
H'_n(x) &= H_{n-1}(x) \nonumber \\
(n+1)H_{n+1}(x) &= xH_n(x)-H_{n-1}(x),\\
H_n(-x) &= (-1)^nH_n(x), \nonumber
\end{align}
for all $n\ge 1$. For $X,Y$ that are $N(0,1)$ random variables that are jointly Gaussian, it can be shown (see, for example, Section 2.10 of \cite{kolokoltsov2011markov}) that
\beq
\langle H_n(X), H_m(Y)\rangle = \E (H_n(X) H_m(Y)) = \left\{ \begin{array}{ll}
	0 & \text{if $m \ne n$,}\\
	\frac{1}{n!} (\E XY)^n & \text{if $m=n$.}
\end{array} \right.
\eeq
Now we would like to expand the function $\tanh(\sigma x/2)$ in terms of orthogonal Hermite polynomials as
\[
\tanh(\sigma x/2) = \sum_{n=0}^\infty C_n H_n(x).
\]
To calculate the coefficients $C_n$, simply note that
\[
C_n = \frac{\big\langle \tanh(\sigma X/2), H_n(X) \big\rangle}{\big\langle H_n(X), H_n(X) \big\rangle} = \frac{(-1)^n}{(2\pi)^{1/2}}\int \tanh\bigg( \frac{\sigma x}{2}\bigg)\frac{d^n}{dx^n}(e^{-x^2/2}) dx.
\]
Denote $\phi(x) = e^{-x^2/2}$, we have
\[
C_n =\frac{ (-1)^n}{\sqrt{2\pi}}\int \tanh\bigg( \frac{\sigma x}{2}\bigg)\phi^{(n)}(x) dx
\]
Note that $\phi(x)$ is an even function and $\tanh(x)$ is an odd function, so the integrand $\phi^{(n)}(x)\tanh(\sigma x/2)$ is an odd function for all odd $n>0$. Therefore $C_{2k} = 0$ for any $k\ge0$. Now we calculate for $k\ge 0$, through integration by parts,
\begin{align*}
C_{2k+1} &=\frac{ (-1)^{2k+1}}{\sqrt{2\pi}}\int \tanh\bigg( \frac{\sigma x}{2}\bigg)\phi^{(2k+1)}(x) dx=\frac{ (-1)^{2k+1}}{\sqrt{2\pi}} \int \tanh^{(1)}\bigg( \frac{\sigma x}{2}\bigg)\phi^{(2k)}(x) dx\\
&=...=\frac{ (-1)^{2k+1}}{\sqrt{2\pi}} \int \tanh^{(2k+1)}\bigg( \frac{\sigma x}{2}\bigg)\phi(x) dx.
\end{align*}
By the fact that, for any $x\ge 0$,
\[
\tanh^{(n)}(x/2)\le \sinh^{(n)}(x),
\]
we have
\begin{align*}
C_{2k+1}&\le \frac{ (-1)^{2k+1}}{\sqrt{2\pi}} \int \sinh^{(2k+1)}(\sigma x )\phi(x) dx\\
&= \frac{ (-1)^{2k+1}}{\sqrt{2\pi}} \int \sinh (\sigma x)\phi^{(2k+1)}(x) dx\\
&=\frac{ 2}{\sqrt{2\pi}}\int_0^\infty \sinh( \sigma x )H_{2k+1}(x)(2k+1)!\phi(x)dx\\
&=\sigma^{2k+1}e^{\sigma^2/2},
\end{align*}
where the last equation follows from Equation 7.387.1 of \cite{gradshteyn2014table}.
As a result, 
\begin{align*}
&\bigg\langle \tanh\bigg(\frac{\sigma X}{2}\bigg), \tanh\bigg(\frac{\sigma Y}{2}\bigg) \bigg\rangle = \bigg\langle \sum_{n=0}^\infty C_n H_n(X),\sum_{n=0}^\infty C_n H_n(Y)\bigg\rangle= \sum_{n=0}^\infty C_n^2 \langle H_n(X), H_n(Y)\rangle \\
&= \sum_{k=0}^\infty \frac{C_{2k+1}^2\rho^{2k+1}}{(2k+1)!}\le e^{\sigma^2}\sum_{k=0}^\infty \frac{(\sigma^2\rho)^{2k+1}}{(2k+1)!}=e^{\sigma^2}\sinh(\sigma^2\rho).
\end{align*}
Now since $\sinh(x) \le 2x$ for $0\le x\le 1$. To see this, note that
\[
\frac{d}{dx} (\sinh(x)-2x) =\cosh(x)-2 \le 0
\]
when $0\le x\le 1$. So $\sinh(x)-2x$ takes its maximum at $x=0$, which is $0$. Thus, given the fact that $\sigma^2 \le1$, we have 
\beq \label{tanh}
\sinh( \sigma^2\rho) \le 6{\sigma^2\rho},
\eeq
which completes the proof.
\qed


\paragraph{Proof of Lemma \ref{lip.lem}.} Since $\ddot{f}(x) = \frac{e^x(1-e^{2x})}{(1+e^x)^4}< \frac{e^x}{(1+e^x)^2}= \dot{f}(x)$ for all $x\in \R$, by mean value theorem, for any $a,b \in \R$, we have for some $c$ between $a$ and $b$,
\[
|\dot{f}(a)-\dot{f}(b)| = |a-b|\ddot{f}(c) <|a-b|\dot{f}(c)\le |a-b| \max\{\dot{f}(a),\dot{f}(b)\}
\]
by monotonicity of $\dot{f}(x)$. The rest of the proof follows from
\[
\bigg|\frac{1}{\dot{f}(a)}-\frac{1}{\dot{f}(b)}\bigg| = \frac{|\dot{f}(a)-\dot{f}(b)|}{\dot{f}(a)\dot{f}(b)}.
\]
\qed

\section{Supplementary Tables and Figures of Section 5.2}

In Section 5.2 of our main paper, we carried our simulations that compare different methods that control FDR. The design covariates were generated from a truncated multivariate Gaussian distribution, whose covariance matrix is a blockwise diagonal matrix of 10 identical unit diagonal Toeplitz matrices as follows
\[
\begin{bmatrix}
1& \frac{p-2}{10(p-1)}&\frac{p-3}{10(p-1)}& ...& \frac{1}{10(p-1)} & 0\\
\frac{p-2}{10(p-1)} & 1 & \frac{p-2}{10(p-1)}&...&\frac{1}{10(p-1)} & \frac{2}{10(p-1)}\\
\vdots &&\ddots&&&\\
0& \frac{1}{10(p-1)} &  \frac{2}{10(p-1)} &...&\frac{p-2}{10(p-1)}&1
\end{bmatrix}.
\]
Due to the space limit, we only presented the boxplots for the pooled empirical FDRs across all the settings. As a complement to Figure 2 in the main paper, the case-by-case empirical FDRs are shown in Figure 1.

\begin{figure}[h!]
	
	\centering
	\includegraphics[angle=0,width=16cm]{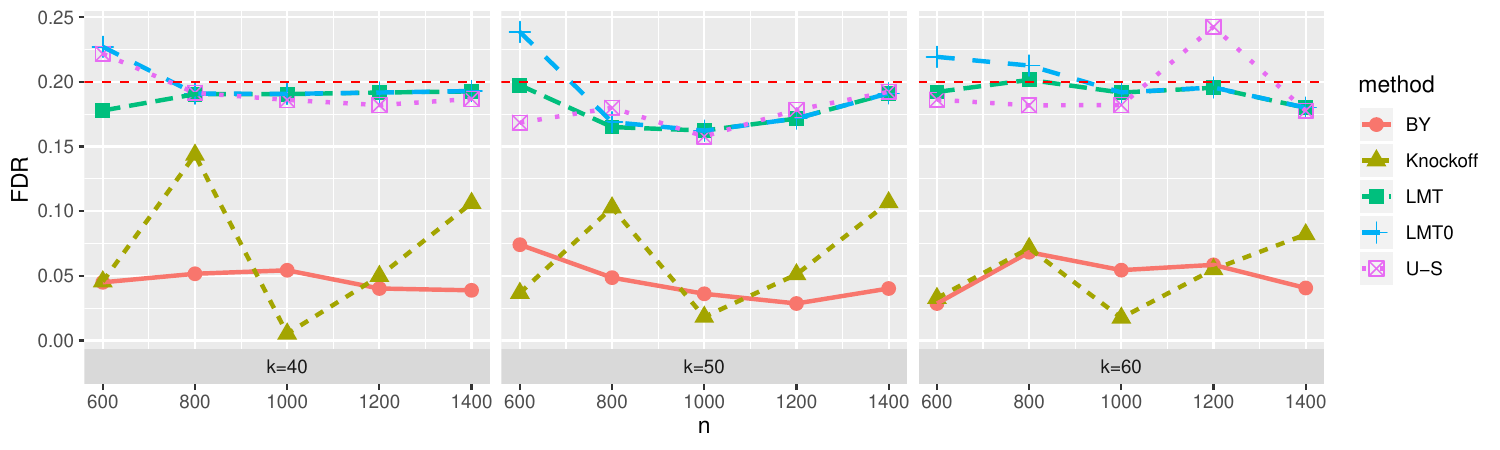}
	\includegraphics[angle=0,width=16cm]{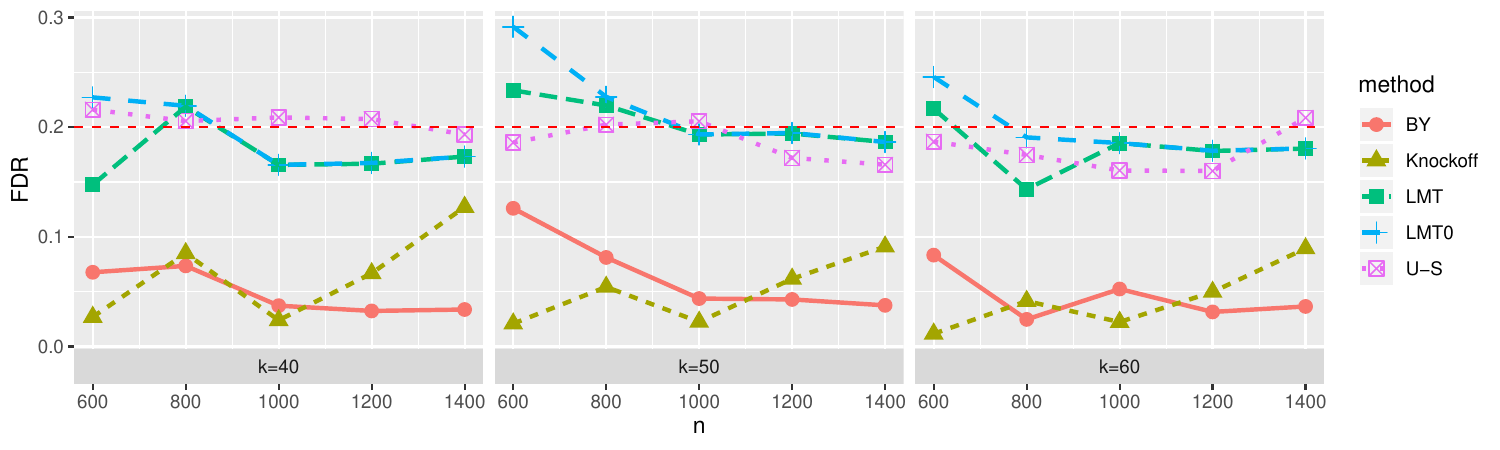}
	\caption{Empirical FDRs under nominal $\alpha=0.2$ for $\rho=3$ (top) and $\rho=4$ (bottom).}\label{pow.res}
\end{figure} 

\appendix

\section{Technical Results for Two-Sample Testing}

In this section, we discuss the implications of our results on single logistic regression problems to the two-sample settings. 
\subsection{Two-Sample Global Hypothesis Testing}

For testing two-sample global null hypothesis
\[
H_0: \beta^{(1)}=\beta^{(2)} \quad \text{vs.} \quad H_1: \beta^{(1)}\ne\beta^{(2)}.
\]
Informed by the previous results, we construct the global two-sample testing procedure as follows.
First we obtain $\check{\beta}_j^{(\ell)}$ and $\tau_j^{(\ell)}$ for each group, and calculate the coordinate-wise standardized statistics 
\[
T_j = \frac{\check{\beta}_j^{(1)}}{\sqrt{2}\tau_j^{(1)}} -\frac{\check{\beta}_j^{(2)}}{\sqrt{2}\tau_j^{(2)}},
\]
for $j=1,...,p$. Then we calculate the difference
the global test statistics is defined as
\beq \label{test.or}
T_n = \max_{1\le j\le p} T_j^2.
\eeq
The following corollary states the asymptotic null distribution for the global test statistics $M_n$ under bounded design. In particular, we assume the parameters $(\beta^{(\ell)},\Sig^{(\ell)})$ for $\ell=1,2$ come from the same parameter space $\Theta(k)$. We denote $\theta=(\beta^{(1)},\Sig^{(1)},\beta^{(2)},\Sig^{(2)}).$
\bet \label{null.dis.two}
Let $T_n$ be the test statistics defined in (\ref{test.or}), $D^{(\ell)}$ be the diagonal of $[\Sig^{(\ell)}]^{-1}$ and $(\xi^{(\ell)}_{ij}) = [D^{(\ell)}]^{-1/2}[\Sig^{(\ell)}]^{-1}[D^{(\ell)}]^{-1/2}$. Suppose $\max_{1\le i<j\le p} |\xi^{(\ell)}_{ij}|\le c_0<1$ for some constant $0<c_0<1$, $\log p = O(n^r)$ for some $0<r<1/5$. and
\begin{enumerate}
	\item under the Gaussian design, we assume (A1) (A3) and $k= o\big( \sqrt{n}/\log^3 p\big)$; or
	\item under the bounded design, we assume (A2) (A3) and $k= o\big( \sqrt{n}/\log^{5/2} p\big).$
\end{enumerate}
Then under $H_0$, for any $x\in \R$, 
Then under $H_0$, for any $x\in \R$, 
\[
P_\theta\big( T_n-2\log p+\log \log p\le x \big) \to \exp\bigg( -\frac{1}{\sqrt{\pi}}\exp (-x/2)\bigg),\quad \text{as $(n,p)\to \infty$}.
\]
\eet

Based on the limiting null distribution, the asymptotically $\alpha$ level tests can be defined as follows:
\[
\Phi_\alpha(T_n)= I \{ T_n \ge 2\log p-\log\log p+q_\alpha\},
\]
where $q_\alpha = -\log(\pi)-2\log\log(1-\alpha)^{-1}.$
The null hypothesis $H_0$ is rejected if and only if $\Phi_\alpha(T_n)=1$.

\subsection{Two-Sample Multiple Hypotheses Testing}

Consider simultaneously testing the two-sample hypothesis
\[
H_{0,j}: \beta_j^{(1)} =\beta_j^{(2)} \quad \text{vs.} \quad H_{1,j}: \beta_j^{(1)}\ne \beta_j^{(2)},\quad j=1,...,p.
\]
As a consequence of the previous analysis, we propose the following two-sample multiple testing procedure controlling FDR/FDP or FDV. 

\paragraph{Two-Sample FDR/FDP Control Procedure.} Define test statistics
\[
T_j = (\check{\beta}^{(1)}_j /\tau^{(1)}_j-\check{\beta}^{(2)}_j /\tau^{(2)}_j)/\sqrt{2},
\]
for $j=1,...,p$.
Let $0<\alpha <1$ and define
\beq \label{t.hat.fdr.two}
\hat{t} = \inf\bigg\{ 0\le t\le b_p: \frac{pG(t)}{\max \big\{\sum_{j=1}^p I\{ |T_{j}| \ge t\},1   \big\}} \le \alpha  \bigg\}.
\eeq
We reject $H_{0,j}$ whenever $|M_{j}| \ge \hat{t}$.

\paragraph{Two-Sample FDV Control Procedure.} For a given tolerable number of falsely discovered variables $r$, let 
\beq
\hat{t}_{FDV} = G^{-1}(r/p).
\eeq 
We reject $H_{0,j}$ whenever $|T_{j}| \ge \hat{t}_{FDV}$.

The following theorems provide the asymptotic behavior of our proposed testing procedures. For simplicity, we only consider the bounded design scenario.

\bet \label{fdr.two}
Assume the conditions of Proposition 1 are satisfied for each group of the samples, we have
\beq
\lim_{(n,p)\rightarrow \infty} \frac{\text{FDR}_\theta(\hat{t})}{\alpha p_0/p} \le 1, \quad\quad \lim_{(n,p)\rightarrow \infty} P_\theta\bigg(\frac{\text{FDP}_\theta(\hat{t})}{\alpha p_0/p}\le 1+\epsilon\bigg) = 1.
\eeq
for any $\epsilon>0$.
\eet

\bet \label{fdv.two}
Suppose the conditions of Theorem 5 are satisfied for each group of the samples, then
\beq
\lim_{(n,p)\rightarrow \infty} \frac{\text{FDV}_\theta(\hat{t}_{FDV})}{rp_0/p} \le 1.
\eeq
\eet

\subsection{Proofs of the Theorems in Appendix A}

In this section, to illustrate how the proofs of the one-sample tests extend to the two-sample tests, we prove Theorem \ref{null.dis.two} in our Appendix A. The proofs of Theorem \ref{fdr.two} and Theorem \ref{fdv.two} are similar and thus are omitted.

\paragraph{Proof of Theorem \ref{null.dis.two}} Define $F^{(\ell)}_{jj}=\E[(\eta_{ij}^{(\ell)})^2/\dot{f}(u_i^{(\ell)})]$ and $n\asymp n_1\asymp n_2$. Define statistics
\[
\tilde{M}_j = \frac{\langle v^{(1)}_j,\epsilon^{(1)}\rangle}{\|v_j^{(1)}\|_n}-\frac{\langle v_j^{(2)},\epsilon^{(2)}\rangle}{\|v_j^{(2)}\|_n}, \quad\quad \check{M}_j = \frac{\sum_{i=1}^{n_1}\eta^{(1)}_{ij}\epsilon_i^{(1)}/\dot{f}(u_i^{(1)})}{\sqrt{n_1F^{(1)}_{jj}}}-\frac{\sum_{i=1}^{n_2}\eta_{ij}^{(2)}\epsilon_i^{(2)}/\dot{f}(u_i^{(2)})}{\sqrt{n_2F^{(2)}_{jj}}}, 
\]
for $j=1,...,p,$ and thereby $\tilde{M}_n = \max_{j}\tilde{M}_j^2, \check{M}_n = \max_{j}\check{M}_j^2$.
\bel \label{event2.two}
Under the condition of Theorem \ref{null.dis.two}, the following events 
\begin{align*}
B_1 = \bigg\{ |\tilde{M}_n-\check{M}_n| =o(1) \bigg\}, \quad B_2 = \bigg\{ |\tilde{M}_n-M_n| =o\bigg( \frac{1}{{\log p}} \bigg) \bigg\},
\end{align*}
hold with probability at least $1-O(p^{-c})$ for some constant $C,c>0$.
\eel
The proof of the above lemma follows directly from the proof of Lemma 1.

It follows that under the event $B_1\cap B_2$, let $y_p = 2\log p-\log \log p+x$ and  $\epsilon_n=o(1)$, we have
\[
P_{\theta}\big( \check{M}_n \le y_p-\epsilon_n \big)\le P_{\theta}\big( {M}_n\le y_p\big)\le P_{\theta}\big( \check{M}_n\le y_p+\epsilon_n \big)
\]
Therefore it suffices to prove that for any $t\in\R$, as $(n,p)\to \infty$,
\beq
P_{\theta}\big( \check{M}_n\le y_p \big) \to  \exp\bigg( -\frac{1}{\sqrt{\pi}}\exp (-x/2)\bigg).
\eeq 
Now define
\[
\hat{M}_j = \frac{\sum_{i=1}^{n_1} \hat{Z}^{(1)}_{ij}}{\sqrt{n_1F^{(1)}_{jj}}}-\frac{\sum_{i=1}^{n_2} \hat{Z}^{(2)}_{ij}}{\sqrt{n_2F^{(2)}_{jj}}} \quad\quad j=1,...,p.
\]
where $\hat{Z}^{(\ell)}_{ij}=v^{0,\ell}_{ij}\epsilon^{(\ell)}_i 1\{ |v^{0,\ell}_{ij}\epsilon^{(\ell)}_i| \le \tau_n\}- \E [v^{0,\ell}_{ij}\epsilon^{(\ell)}_i 1\{ |v^{0,\ell}_{ij}\epsilon^{(\ell)}_i| \le \tau_n\}]$ for $\tau_n = \log p$, $v^{0,\ell}_{ij}= \eta^{(\ell)}_{ij}/\dot{f}(u_i^{(\ell)})$ and $\hat{M}_n = \max_{j}\hat{M}_j^2$. Equivalently, we can write
\[
\hat{M}_j  =  \frac{1}{n_1}\sum_{i=1}^{n_1+n_2}w_{i_j} \quad\quad j=1,...,p.
\]
where
\[
w_{i_j} =  \frac{\hat{Z}^{(1)}_{ij}}{\sqrt{F^{(1)}_{jj}}},\quad\quad \text{for $i=1,...,n_1$,}
\]
and
\[
w_{i_j} =  \sqrt{\frac{n_1}{n_2}}\frac{\hat{Z}^{(2)}_{ij}}{\sqrt{F^{(2)}_{jj}}},\quad\quad \text{for $i=n_1+1,...,n_1+n_2$.}
\]
By similar statement in Lemma 2, it suffices to prove that for any $t\in\R$, as $(n,p)\to \infty$,
\beq \label{25}
P_{\theta}\big( \hat{M}_n\le y_p \big) \to  \exp\bigg( -\frac{1}{\sqrt{\pi}}\exp (-x/2)\bigg).
\eeq 
By Lemma 3 in the main paper, for any integer $0<q<p/2$,
\begin{align} \label{26}
\sum_{d=1}^{2q}(-1)^{d-1} \sum_{1\le j_1 <...<j_d \le p} P_\theta\bigg( \bigcap_{k=1}^d A_{j_k} \bigg) &\le P_\theta \bigg( \max_{1\le j\le p} \hat{M}_j^2 \ge y_p \bigg) \nonumber \\
& \le \sum_{d=1}^{2p-1}(-1)^{d-1}\sum_{1\le j_1 <...<j_d \le p} P_\theta\bigg( \bigcap_{k=1}^d A_{j_k} \bigg),
\end{align}
where $A_{j_k} = \{\hat{M}_{j_k}^2\ge y_p\}$. Now let $\bold{W}_i=(w_{i,j_1},...,w_{i,j_d})^T$ for $1\le i\le n_1+n_2$. Define $\| \bold{a}\|_{\min} = \min_{1\le i\le d}|a_i|$ for any vector $\bold{a}\in \R^d$. Then we have
\[
P_\theta\bigg( \bigcap_{k=1}^d A_{j_k} \bigg) = P_\theta \bigg(\bigg\| n_1^{-1/2}\sum_{i=1}^{n_1+n_2} \bold{W}_i \bigg\|_{\min} \ge y^{1/2}_p  \bigg).
\] 
Then it follows from Theorem 1.1 in \cite{zaitsev1987gaussian} that 
\begin{align}\label{27}
P_\theta \bigg(\bigg\| n_1^{-1/2}\sum_{i=1}^{n_1+n_2} \bold{W}_i \bigg\|_{\min} \ge y^{1/2}_p  \bigg) &\le P_\theta \bigg( \| \bold{N}_d\|_{\min} \ge y_p^{1/2}-\epsilon_n(\log p)^{-1/2} \bigg) \nonumber \\
&\quad+ c_1d^{5/2}\exp \bigg\{-\frac{n_1^{1/2}\epsilon_n}{c_2d^3\tau_n(\log p)^{1/2}}  \bigg\},
\end{align}
where $c_1>0$ and $c_2>0$ are constants, $\epsilon_n \to 0$ which will be specified later, and $\bold{N}_d = (N_{m_1},...,N_{m_d})$ is a normal random vector with $\E(\bold{N}_d) = 0$ and $\text{cov}(\bold{N}_d) = \text{cov}(\bold{W}_1)+n_2/n_1\text{cov}(\bold{W}_{n_1+1})$. Here $d$ is a fixed integer that does not depend on $n,p$. Because $\log p = o(n^{1/5})$, we can let $\epsilon_n \to 0$ sufficiently slowly, say, $\epsilon_n = \sqrt{\frac{\log^5p}{n}}$, so that for any large $c>0$,
\beq\label{28}
c_1 d^{5/2} \exp \bigg\{ -\frac{n^{1/2}\epsilon_n}{c_2 d^3 \tau_n (\log p)^{1/2}}\bigg\} = O(p^{-c}).
\eeq 
Combining (\ref{26}), (\ref{27}) and (\ref{28}), we have
\begin{align} \label{29}
P_\theta \bigg( \max_{1\le j\le p} \hat{M}_j^2 \ge y_p \bigg) \le  \sum_{d=1}^{2p-1}(-1)^{d-1}\sum_{1\le j_1 <...<j_d \le p} P_\theta \bigg( \| \bold{N}_d\|_{\min} \ge y_p^{1/2}-\epsilon_n(\log p)^{-1/2} \bigg) +o(1).
\end{align}
Similarly, one can derive
\beq \label{30}
P_\theta \bigg( \max_{1\le j\le p} \hat{M}_j^2 \ge y_p \bigg) \ge  \sum_{d=1}^{2p}(-1)^{d-1}\sum_{1\le j_1 <...<j_d \le p} P_\theta \bigg( \| \bold{N}_d\|_{\min} \ge y_p^{1/2}+\epsilon_n(\log p)^{-1/2} \bigg) +o(1).
\eeq
Using Lemma 4 in the main paper, it then follows from (\ref{29}) and (\ref{30}) that
\begin{align*}
\limsup_{n,p\to \infty} P_\theta \bigg( \max_{1\le j\le p} \hat{M}_j^2 \ge y_p \bigg) &\le  \sum_{d=1}^{2p}(-1)^{d-1}\frac{1}{d!}\bigg( \frac{1}{\sqrt{\pi}}\exp (-t/2) \bigg)^d, \\
\liminf_{n,p\to \infty} P_\theta \bigg( \max_{1\le j\le p} \hat{M}_j^2 \ge y_p \bigg) &\ge  \sum_{d=1}^{2p-1}(-1)^{d-1}\frac{1}{d!}\bigg( \frac{1}{\sqrt{\pi}}\exp (-t/2) \bigg)^d, 
\end{align*}
for any positive integer $p$. By letting $p\to \infty$, we obtain (\ref{25}) and the proof is complete.
\qed

\label{lastpage}

\end{document}